\documentclass[fleqn,square,numbers,sort&compress,reprint]{elsarticle}

\usepackage{bm}
\usepackage{amsmath}
\usepackage{amssymb}
\usepackage{revsymb}
\usepackage{subcaption}
\usepackage{graphicx}
\usepackage{fourier}
\usepackage{color}
\usepackage[colorlinks=true,linkcolor=blue,citecolor=blue,urlcolor=blue]{hyperref}

\DeclareMathAlphabet{\mathcal}{OMS}{cmsy}{m}{n}
\DeclareMathOperator{\sinc}{sinc}

\begin{document}

\begin{frontmatter}
\title{A General PSTD Method to Solve Quantum Scattering in the Fresnel and Far-Field Regions by A Localized Potential of Arbitrary Form}

\author{Kun Chen}
\ead{kunchen@siom.ac.cn, kunchen@alum.mit.edu}
\affiliation{organization={Key Laboratory of Quantum Optics, Department of Aerospace Laser Technology and Systems,
	Shanghai Institute of Optics and Fine Mechanics, Chinese Academy of Sciences},
   city={Shanghai},
   postcode={201800},
   country={China}}

\begin{abstract}
   We present a time domain method to solve quantum scattering by an arbitrary
   potential of finite range.  The scattering wave function in full space can be
   obtained, including the near field, the mid field (i.e. Fresnel region)
   and the far field.  This is achieved by extending several techniques of FDTD
   computational electrodynamics into the quantum realm.  The
   total-field\slash scattered-field scheme naturally incorporates the incidence source
   condition.  The wave function in the internal model, including the
   interaction region and the close near field, is directly computed through
   PSTD\slash FDTD iterations.  The quantum version of surface equivalence theorem is
   proven and links the wave function in the external free space to the
   PSTD\slash FDTD solution in the internal model.  Parallel implementation of PSTD
   based on overlapping domain decomposition and FFT on local Fourier-basis is
   briefly discussed.  These building blocks unite into a numerical system that
   provides a general, robust solver to potential scattering problems.  Its
   accuracy is verified by the established partial wave method, by comparing the
   predictions of both on the central square potential scattering.  Further
   investigations show the far-field solution is inadequate for simulating
   Fresnel-region effects.
\end{abstract}


\end{frontmatter}


\section{Introduction}
Scattering is a fundamental methodology of physics to detect the structure of
matter and study the interaction between the probing particle and the target.
In the scenario of nonrelativistic matter wave incidence, such as thermal or cold
neutron beam, the problem falls into the category of quantum potential
scattering governed by the Schr\"{o}dinger equation.  Since the establishment of
quantum theory, methods solving scattering problems have been zealously pursued.
In today's standard textbook, rigorous solution is still restricted to the
central potential scattering at the far field limit, where the spherical Bessel
functions asymptotically approach sine and cosine
functions~\cite{book:Sakurai2011,book:Cohen-Tannoudji2005}.  By partial wave
analysis, a phase shift can be numerically retrieved for each partial wave
component, and the total scattering wave function at $r\to\infty$ becomes the
supposition of all these partial waves.  For other forms of potentials,
unfortunately there is no general method to calculate the solution in a strict
way, even in numerical sense.  The scattering wave function may be expressed
in an analytically closed form, such as the Lippmann-Schwinger integral equation,
but it is hard to convert it into actual numbers for comparison with experiment
data.  Often, approximate methods, such as the Born approximation, become the
choice.  Furthermore, all these methods can only give results at the far field.
They are incapable of predicting the scattering behavior in the Fresnel region.

The Fresnel region sits between the near field and the far field.  Its
importance has been realized and explored in the development of a novel imaging
methodology based on the quantum correlation of incident fields, often referred
to as ghost
imaging~\cite{article:HSS-PRL2016,article:Paganin2016,article:Khakimov2016,article:Ratner2018,article:Chen2018,
article:Kingston2020,article:Hodgman2019}.  In this scheme, a spatially
incoherent wave, either an X-ray or a matter wave, is split into a reference arm
and an object arm.  Quantum information is encoded in the correlation between
the two wavefronts.  In the object arm, the detector is placed in the Fresnel
region of the scattered field.  Intensity correlations between the signals of
the two arms are recorded.  As is well known, the far field is subject to
diffraction limit, whereas the near field poses special instrumental
difficulties in arranging the detector.  The transmitted signal collected in the
Fresnel region provides an intermediately close, sub-diffraction-limit
resolution "see-through" of the internals of the target with relatively easy
experimental setup.

The boundary between the Fresnel region and the far field is typical
characterized by a length $D^2/\lambda$, where $D$ is the size of interaction
zone and $\lambda$ the probing wavelength.  In the case of magnetic neutron
scattering, the magnetic induction $\mathbf{B}$ generally scales as $r^{-3}$,
so the effective $D$ will be one order of magnitude larger than the object's
magnetic structure.

Apart from the field distance, another difficulty in solving potential
scattering is the form of the potential function.  A general potential could be
non-central, time-dependent, even nonlinear (i.e. depending on the wave function
itself), and in the event of magnetic neutron scattering, in a matrix form
proportional to the vector product of the Pauli matrices and the magnetic
induction $\mathbf{\sigma}\cdot\mathbf{B}$~\cite{book:Squires1978}.  A precise solution even in the far
field is still out of reach by conventional methods.

Electromagnetic wave scattering and de Broglie wave scattering share many
similarities.  Both have to tackle the problem of solving time-dependent partial
differential equations, with a target setup, a numerical injection of the
incident wave, and the goal of finding the response at distant sites.
After decades' intense development, the finite-difference time-domain (FDTD)
technique and its close variant, the pseudo-spectral time-domain (PSTD)
technique, have become the standard solver of the Maxwell equations governing
electromagnetic scattering~\cite{book:Taflove2005}.  This is the more suitable choice than other
approaches, such as the finite-element method, regarding scattering.  The
advantages of FDTD\slash PSTD include its capability of large scale modeling and the
ability to handle transient response of pulsed incidence.  The last point is
especially enlightening when we consider neutron spallation sources.

By drawing close analogy between the two kinds of scatterings, many concepts
from the FDTD\slash PSTD electrodynamics can be directly extended into the realm of quantum
mechanics.  Among these are total-field\slash scattered-field (TF\slash SF), incident wave
source conditions, surface equivalence theorem, virtual surface phasor
quantities, and near-to-far-field transformation~\cite{book:Taflove2005}.  So far these ideas have been
absent in quantum mechanics.  Of course, the mathematical expressions of their 
quantum version would be distinct from the corresponding electromagnetic version.
In the following sections, we will incorporate the above concepts into the
Schr\"{o}dinger equation, derive their theoretical forms from the first
principle, develop the FDTD\slash PSTD numerical algorithms, and combine the various
aspects into a modern, general, systematic analyzing technique for quantum
potential scattering.  The actual coding and large-scale parallel implementation
will be focused on Fourier PSTD.

\section{Internal Model}
\begin{figure}
   \centering
   \includegraphics[width=0.75\linewidth]{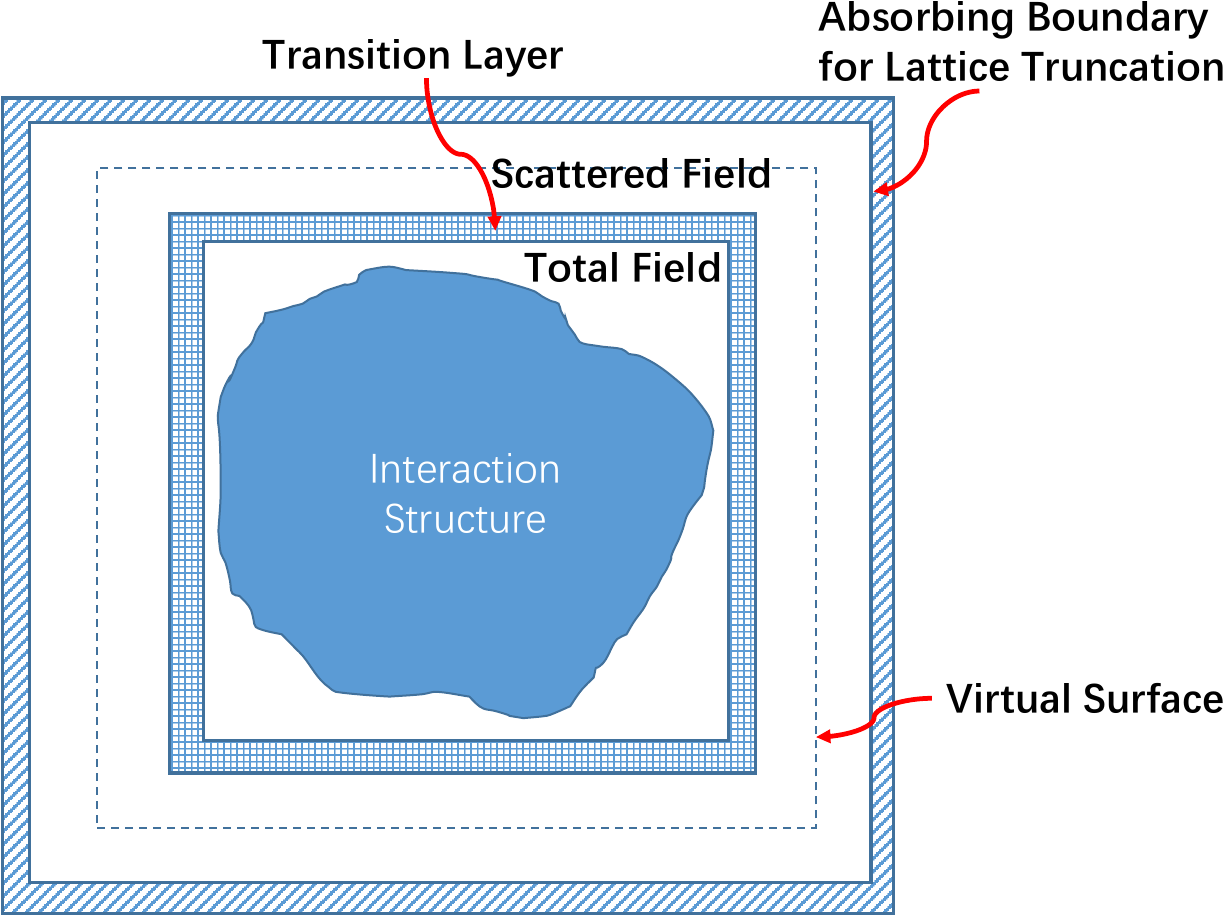}
   \caption{Setup of the Internal Model}
   \label{fig:model-setup}
\end{figure}

The incident wave does not vanish at $\mathbf{r}\to\infty$.  So the scattering
is an unbound state problem of the time-dependent Schr\"{o}dinger equation
(TDSE).  This does not mean TDSE has to be solved in infinite space.  A careful
observation reveals that outside the interaction region, the free propagation of
the scattered wave can be separated as an isolated process.  Thus TDSE only
needs to be solved in a finite space if the interaction is confined to a limited
range.  Consequently, the entire simulation is divided into two stages.  In the
first stage, an internal model is constructed (Fig.~\ref{fig:model-setup}),
consisting of only the interaction region, some nearby free space, and a
truncation boundary.  FDTD/PSTD is applied to the internal model.  In the second
stage, the scattered wave function outside the internal model is calculated via
near-to-distant-field transformation of the first-stage results.

Figure~\ref{fig:model-setup} illustrates the general idea of the internal
model, bearing resemblance to that of FDTD
electrodynamics~\cite{book:Taflove2005}.  Here, the interaction is totally
contained in the TF region, whereas the transition layer and the SF region are
free space.  While the total wave function is employed in the TF, only the pure
scattering wave is considered in the SF.  The incident wave is handled in the
transition layer.  The details of Fig.~\ref{fig:model-setup} are discussed in
the following subsections.

\subsection{Absorption Boundary Conditions}
Discretizing TDSE on finite space requires lattice truncation, otherwise spurious effects
will appear at boundaries of an "open" lattice, either as numerical reflections in
the finite difference method or as wraparound in the Fourier spectral method.  Over the
past decades, various numerical techniques have been developed to absorb the
outgoing waves and eliminate the reflection from the edges, as if the waves have
transparently propagated into the outside infinite space and will never return.

To impose the absorbing boundary condition (ABC), the outer perimeter of the
model domain is surrounded
by an artificial damping layer of finite width (Fig.~\ref{fig:model-setup}).
The most prominent implementations of the damping include exterior complex
scaling (ECS)~\cite{article:McCurdy2002}, smooth exterior scaling
(SES)\cite{article:Moiseyev1998,article:Riss1998}, perfect matched layer
(PML)~\cite{article:Nissen2010,article:Nissen2011}, and complex absorption
potential (CAP)~\cite{article:Kosloff1986,article:Silaev2018}.  The idea behind
ECS and SES is the analytical continuation of the coordinate variable $x$.  For
example, after entering an absorbing boundary from the left, $x$ is continued to a rising
contour into the upper half complex plane, so that the amplitude of the forward
propagating wave $\exp(ikx)$ attenuates along the path.  PML introduces a
different complex coordinate transformation of variable $x$,  expressed as
\begin{equation}
   x\to x+e^{i\gamma}\int_{x_0}^x \sigma(\omega)d\omega,
   \label{eq:PML}
\end{equation}
where $\sigma(\omega)$ is a non-negative function, called the absorption
profile, and $\gamma$ is a constant coordinate stretch parameter.  Accordingly,
the Schr\"{o}dinger equation undergoes variable change, and extra potential
terms appear in the absorbing layer.

However, our numerical experiments on the one dimension free propagation of Gaussian
wave, using both FDTD and PSTD, show the above coordinate-transform based ABCs
are unstable.  In the ECS and SES cases, the wave function quickly blows up near
the starting grid of the contour.  In the PML case, numerical performance is
slightly better, and the wavepacket initially behaves as expected, but
eventually blows up after sufficient iterations.  Further investigations show
these coordinate-transform based ABCs (ECS, SES, PML) all introduce positive
imaginary potential near the entry grids into the boundaries.  (For example, the two
positive spikes of $Im(V_0)$ in Figure~2 of Ref.~\cite{article:Moiseyev1998}.)
As $i\hbar\partial\psi/\partial t = H\psi+V_R\psi+i V_I\psi$, the positive $V_I$
spikes will cause the wave functions at these grids behave like
$\psi_0\exp(V_It)$.  Eventually the exponential growth at these grids will
spread to the entire lattice.

On the other hand, a masking multiplication method, introduced in
Ref.~\cite{article:Kosloff1986}, does render a stable solution.  Here, each time
iteration comprises two steps.  Firstly, the wave function is time-stepped
according to the Schr\"{o}dinger equation,
\begin{equation}
   \psi_{(0)}^{n+1}=\psi^{n-1}-i\frac{2\Delta t}{\hbar}H\psi^{n},\label{eq:premask}
\end{equation}
and secondly, the wave function in the absorbing boundary is manually attenuated by
multiplying a factor, i.e.,
\begin{equation}
   \psi^{n+1}=\left(1-\gamma(\mathbf{r}) \Delta t\right)\psi_{(0)}^{n+1},\label{eq:mask}
\end{equation}
where
\begin{equation}
   \gamma(d)=U_0/\cosh^2(\alpha\,d),
   \label{eq:Poshcl-Teller}
\end{equation}
with $U_0$ a positive constant, $\alpha$ the decay factor, and $d$ the distance
from the outmost surface of the absorbing layer.  Eq.~(\ref{eq:mask}) is
equivalent to
\begin{equation}
   \frac{\partial\psi}{\partial t}=-\gamma(\mathbf{r})\psi,\label{eq:NIP}
\end{equation}
cooresponding to an extra negative imaginary potential (NIP) added to the
Hamiltonian.~\cite{article:Kosloff1986}

Eqs.~(\ref{eq:premask})-(\ref{eq:mask}) can be combined and rewritten as
\begin{equation}
   \psi^{n+1}=(1-\gamma\Delta t)\left(\psi^{n-1}-i\frac{2\Delta
   t}{\hbar}H\psi^{n}\right),
   \label{eq:totalmask}
\end{equation}
or
\begin{equation}
   \psi^{n+1}=e^{-\gamma\Delta t}\left(\psi^{n-1}-i\frac{2\Delta
   t}{\hbar}H\psi^{n}\right).
   \label{eq:expmask}
\end{equation}
Due to the shape of $\gamma$, $\exp(-\gamma\Delta t)$ serves as a mask.  In
actual coding, it can be calculated and stored in computer memory before the
time-stepping loop starts.

Finally, though the Poshcl-Teller shape of NIP (Eq.~(\ref{eq:Poshcl-Teller}))
guarantees a stable ABC, this does not preclude other choices of NIP.
Especially, the potential can be extended to a complex one with
both real and imaginary parts, as far as the imaginary part
is kept non-positive.  Such design can improve numerical
performance~\cite{article:Zhang1998}.

\subsection{Incident Wave Source Conditions}
As discussed in detail in Ref.~\cite{book:Taflove2005}, injecting incident wave
source into the computation space lattice is nontrivial.  Using hard source, or
inserting the incident wave as an initial condition, at each field location in
the space lattice will cause profound problems.  To overcome the
difficulties, a total-field\slash scattered-field (TF\slash SF) technique has
been developed for plane-wave incidence~\cite{book:Taflove2005}.  In this
scheme, the computation lattice is divided into a central core zone (TF)
surrounded by an external zone (SF) (Fig.~\ref{fig:model-setup}).  While the TF
simulates the detailed wave-structure interaction, the SF concerns purely the
scattered wave.  The incident wave is absent in both the TF and SF.  Its
influence only enters the model at the interface between the TF and the SF.
Therefore, the incident wave values are only needed at a tiny fraction of the
space lattice.  This delicate feature not only expedites code execution, but
also saves computer memory.  Most importantly, it allows simulations on both
pulsed and cw wave incidences.

The Maxwell's equations only involve the first-order spatial derivatives, while
the Schr\"{o}dinger equation depends on the second-order ones.  This distinction
makes the TF\slash SF formulation of the latter considerably more complicated
than the former.  In Fig.~\ref{fig:model-setup}, the TF, the transition layer
and the SF are the central framework of the internal model.   Within the SF, an imaginary,
enclosed virtual surface is set for the purpose of calculating wave function at
distant locations outside the internal model.  The TF zone should be large
enough to contain the effective range of the potential $V(\mathbf{r})$.  A
cutoff of $r$ should be valid if $V(\mathbf{r})$ decreases fast enough when $r$
increases.  Consequently, $V(\mathbf{r})$ is taken as 0 outside the TF.  The
thickness of the SF should be kept thin, in order to control the size of the
internal model and improve code efficiency.  Typically two wavelengths should be
enough.  The virtual surface is configured at the center of the SF.

The TF/SF conversion is based on the fact
\begin{equation}
   \psi^\text{total}(\mathbf{r})=\psi^\text{scat}(\mathbf{r})+\psi^\text{inc}(\mathbf{r}),
   \label{eq:psi_total_scat_inc}
\end{equation}
with $\psi^\text{total}$, $\psi^\text{scat}$, and $\psi^\text{inc}$ the total,
scattered, and incident wave function, respectively.  In our numerical model, we
intend to use one unique wave function $\psi(\mathbf{r})$ for the entire
computation domain, so that
\begin{equation}
   \psi(\mathbf{r})=
   \begin{cases}
      \psi^\text{total}(\mathbf{r}) & \mathbf{r}\in \text{TF}\\
      \psi^\text{scat}(\mathbf{r}) & \mathbf{r}\in \text{SF}.
   \end{cases}
   \label{eq:psiTFvsNTF}
\end{equation}
The $\psi(\mathbf{r})$ within the transilation layer will be discussed below.

\subsubsection{FDTD Version of TF\slash SF}
FDTD solvers of the Maxwell's equations commonly utilize the central finite
difference to discretize the first-order spatial derivatives.  TF\slash SF
conversion can be achieved on one single layer of grids from each side.  On the
other hand, the second-order spatial derivatives in the Schr\"{o}dinger equation
require better approximations in order to maintain high numerical accuracy.  In
FDTD, a good choice is the eighth order central finite
difference stencil~\cite{article:Nissen2010}.  For example, in terms of $y$
\begin{equation}
   \left.\frac{\partial^2\psi}{\partial y^2}\right|_{i,j,k}=\frac{1}{\Delta
   y^2}\left[\alpha_0\left.\psi\right|_{i,j,k}+\sum_{\ell=1}^{4}
   \alpha_\ell\left(\left.\psi\right|_{i,j-\ell,k}
   +\left.\psi\right|_{i,j+\ell,k}\right)\right]
   \label{eq:y_stencil}
\end{equation}
with the coefficients
\begin{equation}
   (\alpha_0,\alpha_1,\alpha_2,\alpha_3,\alpha_4)=
   \left(-\frac{205}{72},\frac{8}{5},-\frac{1}{5},\frac{8}{315},
   -\frac{1}{560}\right).
   \label{eq:stencil_alpha}
\end{equation}
The derivative at the central grid requires 4 grid values to the left and 4 grid
values to the right, plus its own value.  This complicates the TF\slash SF
conversion for the Schr\"{o}dinger equation.

\begin{figure}
   \centering
   \includegraphics[width=0.5\linewidth]{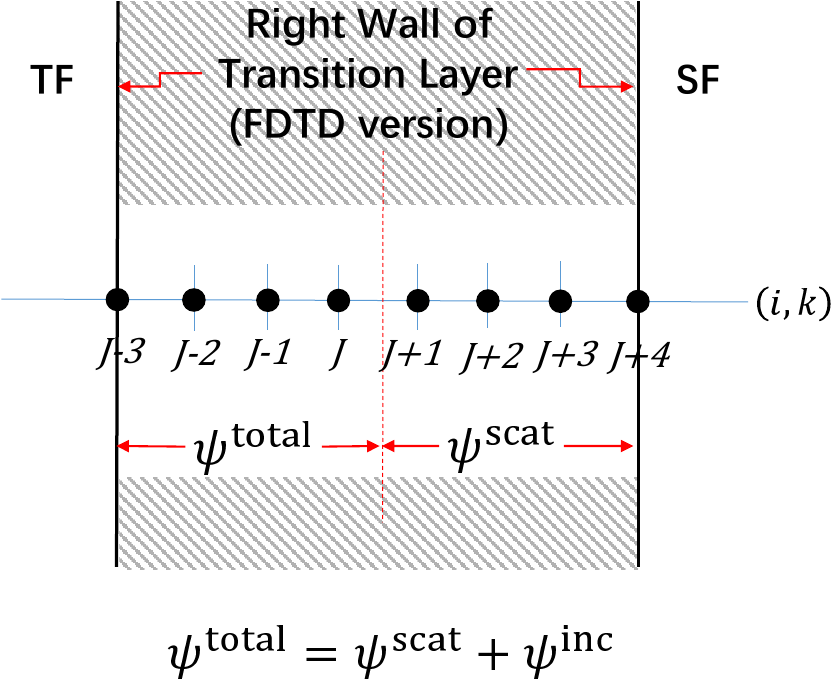}
   \caption{FDTD transition layer and TF\slash SF conversion.  On the four grids
   left to the center line, $\psi=\psi^\text{total}$; on the four grids right to
   the central line, $\psi=\psi^\text{scat}$. }
   \label{fig:FDTD_TF-SF}
\end{figure}

Under central finite difference for the time derivative, the wave function
in the TF and the SF share the same updating formula
\begin{eqnarray}
      \left.\psi\right|^{n+1}_{i,j,k}
      &=&\left.\psi\right|^{n-1}_{i,j,k} -\frac{i2\Delta
      t}{\hbar}V_{i,j,k}+\frac{i\hbar\Delta t}{m}\Biggl\{
      \left(\frac{1}{\Delta x^2}+\frac{1}{\Delta y^2}
      +\frac{1}{\Delta z^2}\right)\alpha_0\left.\psi\right|^n_{i,j,k}\nonumber\\
      &&+\frac{1}{\Delta x^2}\sum_{\ell=1}^4\alpha_\ell\left(\left.
      \psi\right|^n_{i-\ell,j,k}+\left.\psi\right|^n_{i+\ell,j,k}\right)
      +\frac{1}{\Delta y^2}\sum_{\ell=1}^4\alpha_\ell\left(\left.
      \psi\right|^n_{i,j-\ell,k}+\left.\psi\right|^n_{i,j+\ell,k}\right)\nonumber\\
      &&+\frac{1}{\Delta z^2}\sum_{\ell=1}^4\alpha_\ell\left(\left.
      \psi\right|^n_{i,j,k-\ell}+\left.\psi\right|^n_{i,j,k+\ell}\right)
      \Biggr\}.
   \label{eq:FDTDupdatePsi}
\end{eqnarray}
Here, the $\psi$ is the uniq one defined in Eq.~(\ref{eq:psiTFvsNTF}).

What about the grids in the transition layer?  It turns out the transition layer
needs to be 8 grid cells thick, 4 at the TF side and 4 at the SF side.
Figure~\ref{fig:FDTD_TF-SF} illustrates a horizontal $(ik)$ line across the
right wall of the transition layer.  Here, $i$, $j$ and $k$ denote the indices
along the $x$, $y$ and $z$ directions, respectively.  To avoid complication from
the corners, for the time being consider the case where the transverse indices $i$ and $k$ are
within the TF.  Evaluating $\frac{\partial^2\psi}{\partial y^2}$ at $j=J-3,\hdots,J$
would now require grid values from the SF side, and vice versa.

However, both sides of Eq.~(\ref{eq:y_stencil}) must be kept consistent: the
entries must be either all total waves, or all scattered waves.  This is where
the incident wave comes into play.  The consistency condition
Eq.~(\ref{eq:psi_total_scat_inc}) serves to convert a scattered wave at the SF
side to a total wave,
\begin{equation}
   \left.\psi^\text{total}\right|_{i,j,k}=\left.\psi\right|_{i,j,k}
   +\left.\psi^\text{inc}\right|_{i,j,k},\;j\geqslant J+1,
   \label{eq:SFpsi_total}
\end{equation}
and oppositely a total wave at the TF side to a scattered wave, 
\begin{equation}
   \left.\psi^\text{scat}\right|_{i,j,k}=\left.\psi\right|_{i,j,k}
   -\left.\psi^\text{inc}\right|_{i,j,k},\;j\leqslant J.
   \label{eq:TFpsi_scat}
\end{equation}
This connection naturally inserts the plane wave source into the lattice computation.

Substituting Eqs.~(\ref{eq:SFpsi_total}) and (\ref{eq:TFpsi_scat}) into the
Schr\"{o}dinger equation, we can obtain the updating formula for the transition
layer, referred to as the consistency condition.  We list the consistency
condition for the right wall of transition layer as
\begin{eqnarray}
   \left.\psi\right|^{n+1}_{i,J-3,k}
   &=&\left\{\left.\psi\right|^{n+1}_{i,J-3,k}
   \right\}_\text{Eq.~(\ref{eq:FDTDupdatePsi})}+\frac{i\hbar\Delta t}{m\Delta
   y^2}\alpha_4\left.\psi^\text{inc}\right|^n_{i,J+1,k},
   \label{eq:FDTDpsi_extraJminus1}\\
   \left.\psi\right|^{n+1}_{i,J-2,k}
   &=&\left\{\left.\psi\right|^{n+1}_{i,J-2,k}
   \right\}_\text{Eq.~(\ref{eq:FDTDupdatePsi})}+\frac{i\hbar\Delta t}{m\Delta
   y^2}\sum_{\ell=3}^{4}\alpha_\ell\left.\psi^\text{inc}\right|^n_{i,J-2+\ell,k},
   \label{eq:FDTDpsi_extraJminus2}\\
   \left.\psi\right|^{n+1}_{i,J-1,k}
   &=&\left\{\left.\psi\right|^{n+1}_{i,J-1,k}
   \right\}_\text{Eq.~(\ref{eq:FDTDupdatePsi})}+\frac{i\hbar\Delta t}{m\Delta
   y^2}\sum_{\ell=2}^{4}\alpha_\ell\left.\psi^\text{inc}\right|^n_{i,J-1+\ell,k},
   \label{eq:FDTDpsi_extraJminus3}\\
   \left.\psi\right|^{n+1}_{i,J,k}
   &=&\left\{\left.\psi\right|^{n+1}_{i,J,k}
   \right\}_\text{Eq.~(\ref{eq:FDTDupdatePsi})}+\frac{i\hbar\Delta t}{m\Delta
   y^2}\sum_{\ell=1}^{4}\alpha_\ell\left.\psi^\text{inc}\right|^n_{i,J+\ell,k},
   \label{eq:FDTDpsiJ}\\
   \left.\psi\right|^{n+1}_{i,J+1,k}
   &=&\left\{\left.\psi\right|^{n+1}_{i,J+1,k}
   \right\}_\text{Eq.~(\ref{eq:FDTDupdatePsi})}-\frac{i\hbar\Delta t}{m\Delta
   y^2}\sum_{\ell=1}^{4}\alpha_\ell\left.\psi^\text{inc}\right|^n_{i,J+1-\ell,k},
   \label{eq:FDTDpsiJplus1}\\
   \left.\psi\right|^{n+1}_{i,J+2,k}
   &=&\left\{\left.\psi\right|^{n+1}_{i,J+2,k}
   \right\}_\text{Eq.~(\ref{eq:FDTDupdatePsi})}-\frac{i\hbar\Delta t}{m\Delta
   y^2}\sum_{\ell=2}^{4}\alpha_\ell\left.\psi^\text{inc}\right|^n_{i,J+2-\ell,k},
   \label{eq:FDTDpsiJplus2}\\
   \left.\psi\right|^{n+1}_{i,J+3,k}
   &=&\left\{\left.\psi\right|^{n+1}_{i,J+3,k}
   \right\}_\text{Eq.~(\ref{eq:FDTDupdatePsi})}-\frac{i\hbar\Delta t}{m\Delta
   y^2}\sum_{\ell=3}^{4}\alpha_\ell\left.\psi^\text{inc}\right|^n_{i,J+3-\ell,k},
   \label{eq:FDTDpsiJplus3}\\
   \left.\psi\right|^{n+1}_{i,J+4,k}
   &=&\left\{\left.\psi\right|^{n+1}_{i,J+4,k}
   \right\}_\text{Eq.~(\ref{eq:FDTDupdatePsi})}-\frac{i\hbar\Delta t}{m\Delta
   y^2}\alpha_4\left.\psi^\text{inc}\right|^n_{i,J,k}.
   \label{eq:FDTDpsiJplus4}
\end{eqnarray}
The first terms on the r.h.s are given by Eq.~(\ref{eq:FDTDupdatePsi}).
Derivations for the other five walls are straightforward, and their results are
omitted for brevity.

Eqs.~(\ref{eq:FDTDpsi_extraJminus1})-(\ref{eq:FDTDpsiJplus4}) indicate that the
grids of the transition layer undergo a two-step updating, one from the lattice,
and the other from the incident source.  In actual implementation, at each time
leapfrog, all grids in the computation domain are firstly updated according to
Eq.~(\ref{eq:FDTDupdatePsi}), then the grids of the transition layer are further
updated with the incidence source terms, one-by-one for all six walls.  It turns
out this accumulative updating scheme automatically handles the corners of the
transition layer properly, where the walls of different directions overlap.

\subsubsection{PSTD Version of TF\slash SF}
TF\slash SF was not available when PSTD was first introduced into computational
electromagnetics~\cite{article:LiuQH1997}.  The Fourier pseudospectral transform is a global operation.
The transform is performed on all grids (in 1D sense, along $x$, $y$ or $z$).
The discontinuity of the field values across the TF\slash SF interface will
excite spurious ripples in the computation lattice, well known as the Gibbs'
phenomenon.  In addition, the global nature of Fourier transform prevents direct
implementation of
$\mathbf{E}^\text{total}=\mathbf{E}^\text{scat}+\mathbf{E}^\text{inc}$ and
$\mathbf{H}^\text{total}=\mathbf{H}^\text{scat}+\mathbf{H}^\text{inc}$.  Anther
scheme, the pure scattered field formulation becomes an easy choice, in which
the Maxwell's equations are reformulated to express the scattered fields as the
only unknowns on all grids.  However, this approach suffers several drawbacks.
Calculations of the incident wave are needed for all grids and at each time
step, significantly increasing computation burden.  Furthermore, modeling
nonlinear structure becomes much harder, as nonlinear effects depend on total
fields.  This last flaw is especially true to quantum scattering.  The potential
in a nonlinear Schr\"{o}dinger equation is itself a function of the total wave
functions.  The pure scattered field approach is crippled to this kind of
simulation.

\begin{figure}
   \centering
   \includegraphics[width=0.5\linewidth]{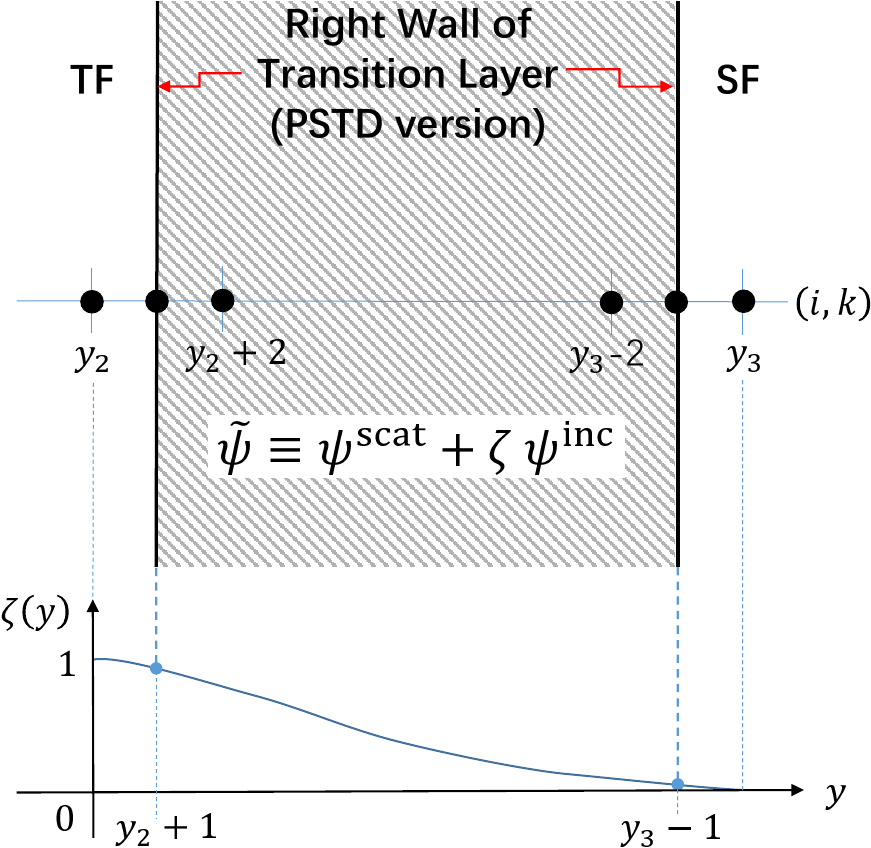}
   \caption{PSTD transition layer and TF\slash SF conversion}
   \label{fig:PSTD_TF-SF}
\end{figure}

TF\slash SF became possible to PSTD when a modified consistency condition was
discovered \cite{article:GaoX2004}.  Adopting the same concept, we can add the
incidence source gradually to the scattered wave across the transition
layer (Fig.~\ref{fig:PSTD_TF-SF}), i.e.,
\begin{equation}
   \tilde{\psi}(\mathbf{r},t)\equiv
   \psi^\text{scat}(\mathbf{r},t)+\zeta(\mathbf{r})\psi^\text{inc}(\mathbf{r},t),
   \label{eq:PSTD_TF-SF}
\end{equation}
where $\zeta$ is the taper function, with value $0$ in the SF and $1$ in the TF, and
rising smoothly from $0$ to $1$ across the transition layer.  Nicely,
$\tilde{\psi}=\psi^\text{total}$ in the TF zone, and $\tilde{\psi}=\psi^\text{scat}$
in the SF zone.  Again, a uniq $\tilde{\psi}$ can be employed in the internal
model.  Furthermore, the smoothness of $\zeta$-function avoids abrupt
changes of $\tilde{\psi}$ and alleviates the Gibbs' oscillation in the PSTD
lattice.

An immediate observation is $\left(\zeta(\mathbf{r})-1\right)V(\mathbf{r})=0$ in the entire
computation domain.  Another important property is
\begin{equation}
   i\hbar\frac{\partial \psi^\text{inc}}{\partial
   t}+\frac{\hbar^2}{2m}\nabla^2\psi^\text{inc}=0
   \label{eq:zeta_psiinc}
\end{equation}
also in the entire computation domain.  Based on these relations, we can obtain
\begin{equation}
	i\hbar\frac{\partial\tilde{\psi}}{\partial
	t}=-\frac{\hbar^2}{2m}\nabla^2\tilde{\psi}
	+V(\mathbf{r})\tilde{\psi}+\frac{\hbar^2}{2m}
	\left(\nabla^2\zeta\,\psi^\text{inc}+2\nabla\zeta\cdot\nabla\psi^\text{inc}\right).
	\label{eq:tildeSchrodinger}
\end{equation}
The last two terms on the r.h.s originate from the incidence wave.  Because
$\nabla^2\zeta$ and $\nabla\zeta$ equal to $0$ in both the TF zone and the SF
zone, these terms only exist in the transition layer.

In 3D, $\zeta(x,y,z)=\zeta_x(x)\zeta_y(y)\zeta_z(z)$ where the three components
share the same function form.  For example,
\begin{equation}
   \zeta_y(y)=
   \begin{cases}
      0	& y\leqslant y_0 \\
      \xi\left(\frac{y-y_0}{y_1-y_0}\right)	& y_0<y<y_1 \\
      1	& y_1\leqslant y\leqslant y_2 \\
      1-\xi\left(\frac{y-y_2}{y_3-y_2}\right)	& y_2<y<y_3	\\
      0	& y\geqslant y_3,
   \end{cases}
   \label{eq:zeta_y}
\end{equation}
where the left transition wall is from $y_0+1$ to $y_1-1$, the right one from $y_2+1$
to $y_3-1$.  Quenching of the Gibbs' phenomenon
depends on a wise choice of the $\xi$-function.  PSTD electrodynamics takes the
integral form of the Blackman-Harris window function (IBH) as the taper
function~\cite{article:GaoX2004}.  However, IBH is not analytical and its
coefficients are empirical.  The precision of IBH is limited to
$10^{-4}$ and in addition, its first derivative is not exactly $0$ at $\rho=1$.
In the Schr\"{o}dinger equation, this tiny mismatch can excite Gibbs'
oscillation to some extent.  For higher precision, we design an optimal
taper,
\begin{equation}
   \xi(\rho)=\rho-\frac{2}{3\pi}\sin(2\pi\rho)+\frac{1}{12\pi}\sin(4\pi\rho),\quad
   0\leqslant\rho\leqslant 1.
   \label{eq:xi}
\end{equation}
Note this function is totally analytical.  Its curve is close to IBH.  Though
its values are not strictly between 0 and 1, only the smoothness matters.  Its
derivates up to the fourth-order are all exactly $0$ at $\rho=0$ and $\rho=1$.
So the curves $\zeta_x$, $\zeta_y$ and $\zeta_z$ are fourth-order smooth.

\subsubsection{Calculation of the Incident Wave}
\begin{figure}
   \centering
   \includegraphics[width=0.7\linewidth]{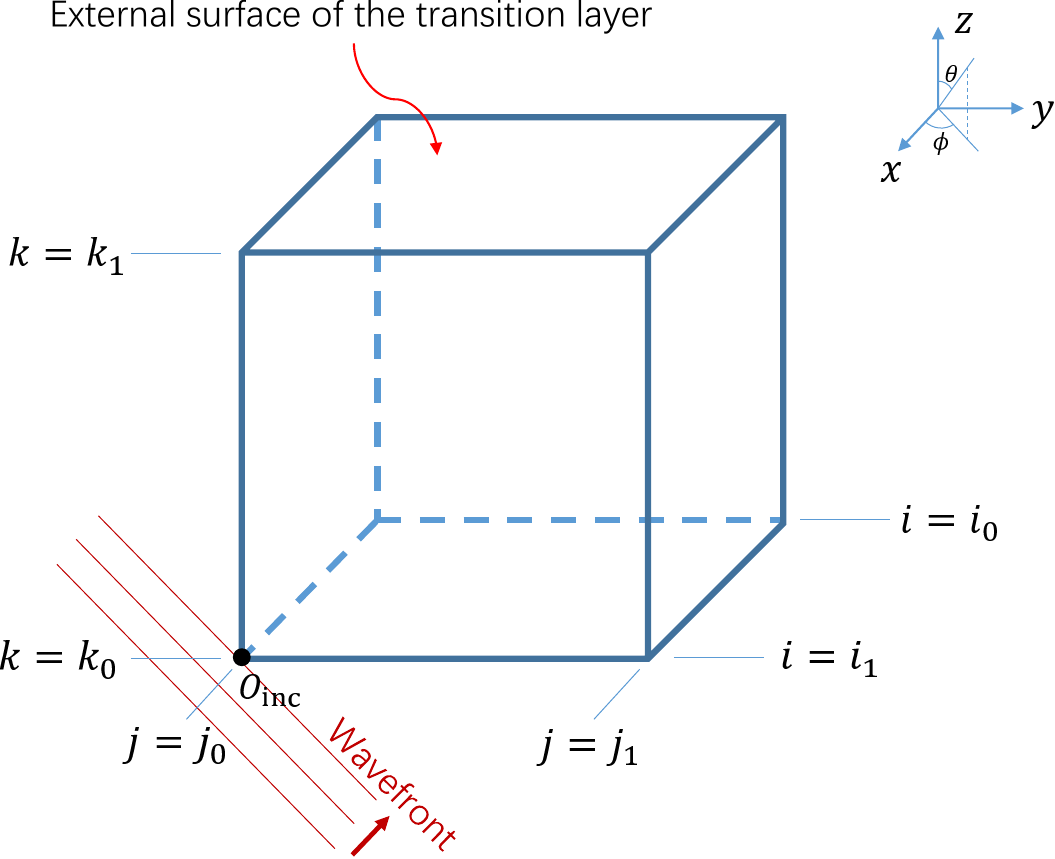}
   \caption{Identification of the origin of the 1D incident wave.}
   \label{fig:incident-source}
\end{figure}

Fig.~\ref{fig:incident-source} illustrates the external surface of the
transition layer.  The wavefront of an incident plane wave will make initial
contact with one of the eight corners.  The source distribution in the
transition layer is a 1D problem, as the grids sitting on the same wavefront
share the same source value.  In only a few cases there exist analytical
expressions of the incident wave functions, such as the sinusoidal wave and the
Gaussian wave packet.  For wave packet of arbitrary shape, the 1D
Schr\"{o}dinger equation for the incident wave has to be solved \textit{in free
space} concurrently with the 3D Schr\"{o}dinger equation of scattering at hand.
At each time iteration, the 1D source is updated first, deployed to the transition
layer, and then the 3D internal model (Fig.~\ref{fig:model-setup}) updated.

The coordinate origin $O_\text{inc}$ of this 1D axis would be the initial
contact point of the incident wave front with the transition layer.  Let the
unit incident wavevector be
\begin{equation}
   \mathbf{\hat{k}}_\text{inc}=\mathbf{\hat{x}}\,\sin\theta\,\cos\phi
   +\mathbf{\hat{y}}\,\sin\theta\,\sin\phi+\mathbf{\hat{z}}\,\cos\theta.
   \label{eq:khat}
\end{equation}
The grid indices $(O_x,O_y,O_z)$ of the initial contact corner is
\begin{equation}
   (O_x,O_y,O_z)=
     \begin{cases}
	(i_0,j_0,k_0) & 0^\circ\leqslant\theta\leqslant 90^\circ,
	0^\circ\leqslant\phi\leqslant
	90^\circ\\
	(i_0,j_1,k_0) & 0^\circ\leqslant\theta\leqslant 90^\circ,
	90^\circ<\phi\leqslant
	180^\circ\\
	(i_1,j_1,k_0) & 0^\circ\leqslant\theta\leqslant 90^\circ,
	180^\circ<\phi\leqslant
	270^\circ\\
	(i_1,j_0,k_0) & 0^\circ\leqslant\theta\leqslant 90^\circ, 270^\circ<\phi<
	360^\circ\\
	(i_0,j_0,k_1) & 90^\circ<\theta\leqslant 180^\circ,
	0^\circ\leqslant\phi\leqslant
	90^\circ\\
	(i_0,j_1,k_1) & 90^\circ<\theta\leqslant 180^\circ,
	90^\circ<\phi\leqslant
	180^\circ\\
	(i_1,j_1,k_1) & 90^\circ<\theta\leqslant 180^\circ,
	180^\circ<\phi\leqslant
	270^\circ\\
	(i_1,j_0,k_1) & 90^\circ<\theta\leqslant 180^\circ, 270^\circ<\phi<
	360^\circ
     \end{cases}
     \label{eq:Oinc}
\end{equation}

The coordinate of grid $(i,j,k)$ on the 1D axis is the projection
\begin{equation}
   d=\left\{(i-O_x)\Delta x\,\mathbf{\hat{x}}+(j-O_y)\Delta
   y\,\mathbf{\hat{y}}+(k-O_z)\Delta
   z\,\mathbf{\hat{z}}\right\}\cdot\mathbf{\hat{k}}_\text{inc}
   \label{eq:inc_d}
\end{equation}
Generally $d$ may not fall exactly on a grid on the 1D axis.  Wave value on
$d$ is often obtained by nearest neighbor interpolation, or by an inverse FFT
method.  Note the incidence source terms on the r.h.s. of
Eq.~(\ref{eq:tildeSchrodinger}) also include derivative of the incident wave.
So the calculations come in pairs,
\begin{eqnarray}
   \left.\psi_\text{inc}^n\right|_{i,j,k}
   &=&\frac{1}{N}\sum_{\ell=-N/2}^{N/2-1}\left.\mathcal{F}[\psi_\text{inc}^n]\right|_\ell
   e^{i\frac{2\pi\ell}{N\Delta}d},
   \label{eq:pstd_psiinc}\\
   \left.\left(\nabla\psi_\text{inc}^n\right)\right|_{i,j,k}
   &=&\frac{\mathbf{\hat{k}}_\text{inc}}{N}\sum_{\ell=-N/2}^{N/2-1}\;i\frac{2\pi
   \ell}{N\Delta}\left. \mathcal{F}\left[\psi_\text{inc}^n\right]\right|_\ell
   e^{i\frac{2\pi\ell}{N\Delta}d},
   \label{eq:pstd_dpsiinc}
\end{eqnarray}
with $N$ the total number of grids along the 1D axis, $\Delta$ the grid size,
$n$ the time step index, and $\mathcal{F}$ the 1D FFT operation.

\subsection{Stability Condition}
The choice of time-marching increment $\Delta t$ is predetermined by the
requirement of stability.  Though a large $\Delta t$ makes the simulation take
fewer iterations to finish, a runaway result is meaningless.  On the other hand,
a small $\Delta t$ increases the computation cost.  An optimal $\Delta t$ should
be large enough, and yet keep the total error bounded at any time-step.

The relationship between $\Delta t$ and the spatial discretization has been
studied in the FDTD-Q scheme~\cite{article:Soriano2004,article:Dai2005}.  In
FDTD-Q the real part and imaginary part of the wave function is time-marched
alternately to avoid complex numerics.  The FDTD-Q stability condition is
inapplicable to this work, since complex computation is preferred in order to
facilitate the FFT and inverse FFT operations.  To derive our version of
stability condition, we adopt the approach of Ref.~\cite{article:Soriano2004}.
The discrete form of Schr\"{o}dinger equation is first separated into a temporal
eigenvalue problem and a spatial eigenvalue problem, i.e.
\begin{eqnarray}
   i\hbar\frac{\psi^{n+1}-\psi^{n-1}}{2\Delta
   t}&=&\lambda_t\psi^n,\label{eq:tempeigen}\\
   -\frac{\hbar^2}{2m}\nabla^2\psi(\mathbf{r},t)+V(\mathbf{r})
   \psi(\mathbf{r},t)&=&\lambda_s\psi(\mathbf{r},t).
   \label{eq:spatialeigen}
\end{eqnarray}
A "growth factor" $q$ is defined to characterize the growth of wave function
during the time iteration, i.e.,
\begin{eqnarray}
   q&=&\psi^{n+1}/\psi^n,\label{eq:q1}\\
   q&=&\psi^{n}/\psi^{n-1}.\label{eq:q2}
\end{eqnarray}
Substituting Eqs.~(\ref{eq:q1})-(\ref{eq:q2}) into Eq.~(\ref{eq:tempeigen}) results in an
equation for $q$,
\begin{equation}
   \frac{1}{2}\left(q-\frac{1}{q}\right)=-i\frac{\lambda_t\Delta
   t}{\hbar}.\label{eq:q_equation}
\end{equation}
Redefine $q=iQ$ and $w=-\lambda_t\Delta t/\hbar$.  Eq.~(\ref{eq:q_equation}) is
essentially the Joukowsky transform
\begin{equation}
   w=\frac{1}{2}\left(Q+\frac{1}{Q}\right).\label{eq:Joukowsky}
\end{equation}

The branch cut for the Joukowsky transform is the unit circle $|Q|=1$.  A $Q$
within the circle and its reciprocal $1/Q$ outside the circle will map to the
same $w$ on the complex plane.  The transform maps a circle in the $Q$-plane to
a ellipse in the $w$-plane, with the foci at 1 and -1 on the real axis.  As
$|Q|$ approaches 1, the ellipse shrinks to the line interval $[-1,1]$ on the real
axis.  On the unit circle, $Q=e^{i\varphi}$ ($0\leqslant\varphi<2\pi$).  There
is $w=\cos\varphi$ and $-1\leqslant w\leqslant 1$.  The stability requirement
imposes condition on $|q|$.  If $|q|\to 0$, $\psi$ vanishes, whereas if
$|q|\to\infty$, $\psi$ explodes.  Neither is stable.  When $|q|$ ($=|Q|$) is
around 1, small $\Delta t$ brings small change in $\psi$.  In this case the
mapped $w$ is in the vicinity of the interval line discussed above.  Thus, we
have the condition for $\lambda_t$
\begin{equation}
   \bigl|Re(w)\bigr|=\Biggl|Re\left(\frac{\lambda_t\Delta
   t}{\hbar}\right)\Biggr|\leqslant 1.\label{eq:lambda_t}
\end{equation}

\subsubsection{FDTD}
To process Eq.~(\ref{eq:spatialeigen}), the most general solution
can be considered as a superposition of plane waves.  For each plane
wave component, the 8th-order central finite difference
(Eqs.~(\ref{eq:y_stencil})-(\ref{eq:stencil_alpha})) is applied to
Eq.~(\ref{eq:spatialeigen}) with $\psi(\mathbf{r},t)=\exp[i(k_x x+k_y y+k_z z-\omega
t)]$.  A calculation gives
\begin{eqnarray}
   \lambda_s&=&\frac{\hbar^2}{2m}\left[\frac{1}{\Delta_x^2}f\left(k_x\Delta_x\right)
   +\frac{1}{\Delta_y^2}f\left(k_y\Delta_y\right)+\frac{1}{\Delta_z^2}f\left(k_z\Delta_z\right)
   \right]+V,\label{eq:spatialFDTDeigen}\\
   f(\delta)&=&\frac{16}{35}\sin^8\frac{\delta}{2}
   +\frac{32}{45}\sin^6\frac{\delta}{2}+\frac{4}{3}\sin^4\frac{\delta}{2}
   +4\sin^2\frac{\delta}{2}.\label{eq:spatialFDTDstencilfactor}
\end{eqnarray}
Because consistency requires $\lambda_t=\lambda_s$, Eq.~(\ref{eq:lambda_t}) is
linked to Eq.~(\ref{eq:spatialFDTDeigen}).  If the maximum value
$\left|Re(w)\right|_{max}\leqslant 1$, Eq.~(\ref{eq:lambda_t}) is guaranteed.  
This leads to the sufficient (but not the necessary) stability condition for the
eighth-order central FDTD
\begin{equation}
   \Delta t\leqslant\frac{\hbar}{\frac{1024\hbar^2}{315m}\left[\frac{1}{\Delta
   x^2}+\frac{1}{\Delta y^2}+\frac{1}{\Delta z^2}\right]+\left|V\right|_\text{max}}.
   \label{eq:FDTDStability}
\end{equation}

\subsubsection{PSTD}
The plane wave supposition argument is unnecessary for PSTD.  In FFT language,
Eq.~(\ref{eq:spatialeigen}) converts to
\begin{equation}
   \lambda_s\psi=-\frac{\hbar^2}{2m}\biggl\{\mathcal{F}_x^{-1}
   \left[-k_x^2 \mathcal{F}_x[\psi]\right]+\mathcal{F}_y^{-1}
   \left[-k_y^2 \mathcal{F}_y[\psi]\right]\\
   +\mathcal{F}_z^{-1}\left[-k_z^2 \mathcal{F}_z[\psi]\right]\biggr\}
   +V\psi,
   \label{eq:spatialFFTeigen}
\end{equation}
where $\mathcal{F}_x$ and $\mathcal{F}^{-1}_x$ denote the 1D FFT and inverse FFT
over the $x$ direction, and so on.
One important property of PSTD is that the kinetic energy representable by
the discrete lattice has an upper bound.  The maximum $k$s on the r.h.s of
Eq.~(\ref{eq:spatialFFTeigen}) are $\pm\pi/\Delta x$, $\pm\pi/\Delta y$ and
$\pm\pi/\Delta z$ respectively.  Any larger $k$ will be aliased to
a value within the limits. Therefore,
\begin{equation}
   \lambda_s<\frac{\hbar^2\pi^2}{2m}\left(\frac{1}{\Delta x^2}+\frac{1}{\Delta
   y^2}+\frac{1}{\Delta z^2}\right)+\left|V\right|_\text{max}. \label{eq:lambdas_limit}
\end{equation}
Again, the consistency requirement $\lambda_t=\lambda_s$ and the condition
$\left|Re(w)\right|_{max}\leqslant 1$ lead to the sufficient (but not the
necessary) stability condition of PSTD
\begin{equation}
   \Delta t\leqslant\frac{\hbar}{\frac{\hbar^2\pi^2}{2m}\left[\frac{1}{\Delta
   x^2}+\frac{1}{\Delta y^2}+\frac{1}{\Delta
   z^2}\right]+\left|V\right|_\text{max}}.\label{eq:PSTDstability}
\end{equation}

Comparing Eq.~(\ref{eq:PSTDstability}) with Eq.~(\ref{eq:FDTDStability}), we
notice that the \textit{critical time step}, the maximum time increment which
maintains the stability, for PSTD is smaller than that for FDTD.

\subsection{PSTD: Elimination of Numerical Phase Velocity Error For Monochromatic Wave}
Deviation of the phase velocity of numerical wave from the true physical
velocity arises when the spatial and time coordinates are discretized.  For
FDTD on a rectangular grid lattice, a numerical wave would
propagate faster along the diagonal of a grid cell than along the three edges.
This anisotropy is absent in PSTD, because the spatial derivatives are converted
into strict spectral operations without resort to finite difference
approximations.  On the other hand, the finite time-step $\Delta t$ still
affects the velocity.  For a monochromatic wave 
$\psi=\exp(i\mathbf{k}\cdot\mathbf{x}-iEt/\hbar)$, the PSTD Schr\"{o}dinger
equation is
\begin{equation}
   -\frac{\hbar^2}{2m}\left\{\mathcal{F}_x^{-1}\left[-k_x^2\mathcal{F}_x\left[\psi\right]\right]
   +\mathcal{F}_y^{-1}\left[-k_y^2\mathcal{F}_y\left[\psi\right]\right]
   +\mathcal{F}_z^{-1}\left[-k_z^2\mathcal{F}_z\left[\psi\right]\right]\right\}
   =i\hbar\frac{\partial\psi}{\partial t}.\label{eq:numdis1}
\end{equation}
Here, $k^2_x$, $k^2_y$, and $k^2_z$ are fixed values, and can be taken out of
the inverse FFTs.  Applying the central finite difference approximation to the
time derivative, we have
\begin{equation}
   k^2=\frac{2mE}{\hbar^2}\sinc\left(\frac{E\Delta t}{\hbar}\right).
   \label{eq:numdis2}
\end{equation}
At the limit $\Delta t\to 0$, Eq.~(\ref{eq:numdis2}) reduces to the physical
formula $\hbar^2k^2=2mE$.  The time discretization distorted the relation
by a factor $\eta=\sinc(E\Delta t/\hbar)$.  If we rescale the kinetic energy
term by factor $\eta$, i.e.
$-\frac{\hbar^2}{2m}\nabla^2\,\to\,-\eta\frac{\hbar^2}{2m}\nabla^2$, this
numerical artifact will completely disappear.  Thus, for monochromatic wave
incidence, the phase velocity discrepancy can be perfectly corrected.
Furthermore, impulsive wave incidence has potentially a broad spectrum.  The
numerical dispersion of phase velocity is cancelled at the central wavelength
and largely reduced at the side wings.

\subsection{PSTD Updating Formula}
A careful examination of Eq.~(\ref{eq:tildeSchrodinger}) reveals that it can be
converted to a dimensionless form.  Assuming the central wavelength of the
incident wave is $\lambda_0$.  Correspondingly other central parameters are:
central wavevector $k_0=2\pi/\lambda_0$, central energy $E_0=\hbar^2k_0^2/2m$,
central reduced wavelength $\lambdabar_0=\lambda_0/2\pi$, and central angular
frequency $\omega_0=E_0/\hbar$.  Define the dimensionless time and spatial
variables as
\begin{equation}
   \tau\equiv\omega_0 t,\quad\bar{x}\equiv x/\lambdabar_0,\quad\bar{y}\equiv
   y/\lambdabar_0,\quad\bar{z}\equiv z/\lambdabar_0.
   \label{eq:rescaled_var}
\end{equation}
Thus one period of time and space are both $2\pi$.  The expression of
Eq.~(\ref{eq:tildeSchrodinger}) is now simplified to
\begin{equation}
   \frac{\partial\tilde{\psi}}{\partial\tau}=i\eta\bar{\nabla}^2\tilde{\psi}
   -i\frac{V}{E_0}\tilde{\psi}-i\left[\bar{\nabla}^2\zeta\;\psi_\text{inc}
   +2\bar{\nabla}\zeta\cdot\bar{\nabla}\psi_\text{inc}\right].
   \label{eq:dimlesstildeSchrodinger}
\end{equation}
Here, the symbol $\bar{\nabla}$ denotes spatial gradient over $\bar{x}$,
$\bar{y}$ and $\bar{z}$.  Further, the stability condition
Eq.~(\ref{eq:PSTDstability}) becomes
\begin{equation}
   \Delta\tau\leqslant\left[\pi^2\left(\frac{1}{\Delta\bar{x}^2}
   +\frac{1}{\Delta\bar{y}^2}+\frac{1}{\Delta\bar{z}^2}\right)
   +\frac{\left|V\right|_\text{max}}{E_0}\right]^{-1}.  \label{eq:dtau_stability}
\end{equation}

We use Eq.~(\ref{eq:expmask}) to impose the ABC. In 3D three independent $\gamma$ functions
(Eq.~(\ref{eq:Poshcl-Teller})), $\gamma(\bar{x})$, $\gamma(\bar{y})$ and
$\gamma(\bar{z})$ are set for the $x$, $y$ and $z$ directions.  The ABC
multiplier
\begin{equation}
   \Gamma(\bar{x},\bar{y},\bar{z})=e^{-\gamma(\bar{x})\,\Delta\tau}\,
   e^{-\gamma(\bar{y})\,\Delta\tau}\,e^{-\gamma(\bar{z})\,\Delta\tau}
   \label{eq:Gamma}
\end{equation}
can be prepared beforehand and stored in computer memory.  The PSTD updating
formula for Eq.~(\ref{eq:dimlesstildeSchrodinger}) is then
\begin{eqnarray}
   \left.\tilde{\psi}\right|^{n+1}_{i,j,k}
   &=&\Gamma_{i,j,k}\Biggl\{
      \left.\tilde{\psi}\right|^{n-1}_{i,j,k}-2i\eta\Delta\tau\biggl(
      \mathcal{F}^{-1}_{\bar{x}}\left[\bar{k}^2_{\bar{x}}
      \mathcal{F}_{\bar{x}}\left[\tilde{\psi}^n\right]\right]\nonumber\\
      &&+\mathcal{F}^{-1}_{\bar{y}}\left[\bar{k}^2_{\bar{y}}
      \mathcal{F}_{\bar{y}}\left[\tilde{\psi}^n\right]\right]
      +\mathcal{F}^{-1}_{\bar{z}}\left[\bar{k}^2_{\bar{z}}
      \mathcal{F}_{\bar{z}}\left[\tilde{\psi}^n\right]\right]\biggr)
      \biggr|_{i,j,k}\nonumber\\
      &&-2i\Delta\tau\frac{V_{i,j,k}}{E_0}\left.\tilde{\psi}\right|
      ^n_{i,j,k}-2i\Delta\tau\left.\left[\bar{\nabla}^2
      \zeta\;\psi_\text{inc}^n+2\bar{\nabla}\zeta\cdot\bar{\nabla}
      \psi_\text{inc}^n\right]\right|_{i,j,k}\Biggr\}.
   \label{eq:PSTDupdating}
\end{eqnarray}
Note almost all FFT software libraries shift the index for $k$ to $0,\hdots,N-1$.
In Eq.~(\ref{eq:PSTDupdating}), when calculating the factor $\bar{k}^2$ within
the $\mathcal{F}^{-1}$, the index must be shifted back to the range
$-N/2,\hdots,N/2-1$.

In the special case of sinusoidal wave incidence, the last term on the r.h.s. of
Eq.~(\ref{eq:PSTDupdating}) can be factorized to a product of phasor component
and a time exponent $\exp(-in\Delta\tau)$.  There is no need to solve a 1D
incidence source equation.  The phasor component can be calculated beforehand
and stored in computer memory.  Calculation of the time exponent has one further
simplification.  Factors $\sin\Delta\tau$ and $\cos\Delta\tau$ can be calculated
beforehand and stored in computer memory.  Since $\sin(n+1)\Delta\tau=\sin
n\Delta\tau\,\cos\Delta\tau+\cos n\Delta\tau\,\sin\Delta\tau$ and
$\cos(n+1)\Delta\tau=\cos n\Delta\tau\,\cos\Delta\tau-\sin
n\Delta\tau\,\sin\Delta\tau$, the sine and cosine of $(n+1)\Delta\tau$ can be
obtained by four multiplications and two additions, where $\sin n\Delta\tau$ and
$\cos n\Delta\tau$ is calculated in time step $n$, stored in computer
memory, recursively used in time step $n+1$, and so on.  This starts at $n=1$.
Thus the costly series call to $\sin$ and $\cos$ functions are avoided.

\subsection{Parallel Implementation of PSTD}
Traditional FFT is a global operation involving all data points across the
entire computation domain.  Modern supercomputers are often distributed-memory
cluster systems interconnected via ultra-fast networks.  Data exchange between
processing units is through message-passing interface (MPI).  The particular
difficulty of PSTD on global basis is the misalignment of FFT and the subsequent
inverse FFT on the same node.  Two massive global data transpositions are
required for each time step, through non-blocking MPI all-to-all data exchange
between all possible node-to-node pairs.  The mutual MPI calls can dramatically
slow down the code running.  Because CPU speeds are orders of magnitude faster
than that of networking units, a high efficiency simulation should promote the
percentage of in-node computation and lower the percentage of inter-node
communications.  The FFT on local-Fourier-basis eliminates the requirement for
all-to-all data exchanges, while retaining the accuracy of the global
FFT~\cite{article:Israeli1993,article:Chen2010}.  This approach contains several
ingredients:

\begin{figure}
   \centering
   \includegraphics[width=0.9\linewidth]{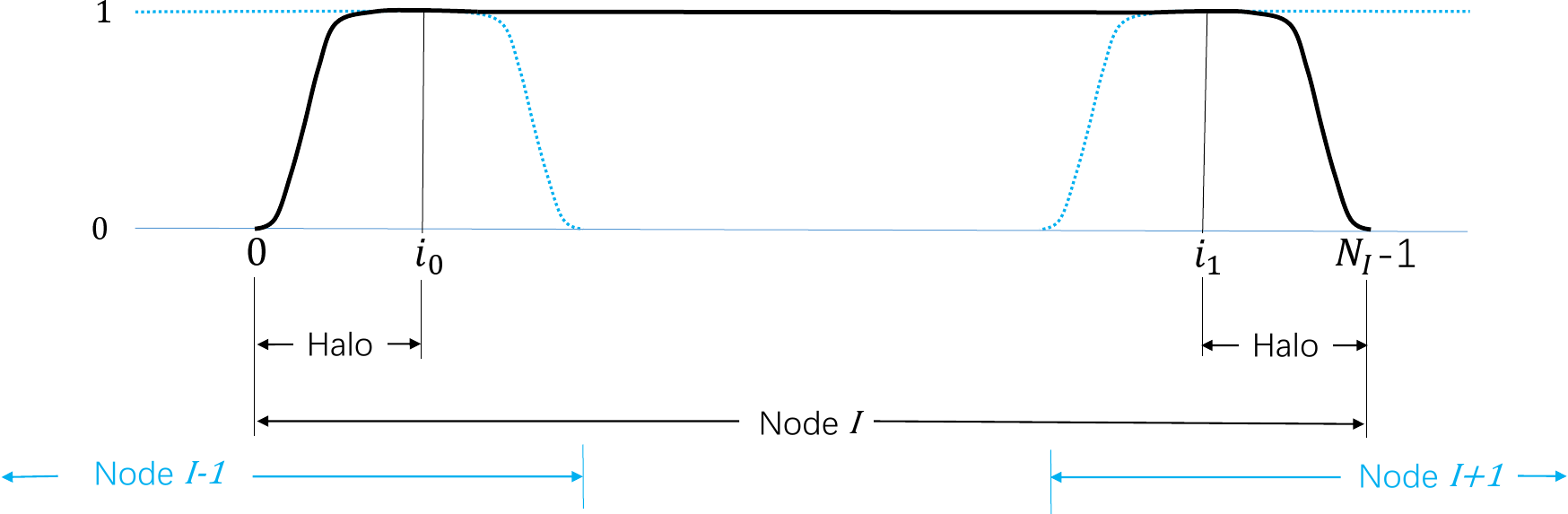}
   \caption{The overlapping between adjacent subdomains and the halo regions.}
   \label{fig:overlapping}
\end{figure}

\begin{itemize}
   \item Overlapping domain decomposition to divided the computation domain into
      sub-domains in the $x$\nobreakdash-, $y$\nobreakdash-, and
      $z$\nobreakdash-directions.  Each sub-domain has halo regions on all six
      domain walls.  Data in the halos are first copied from the corresponding
      grids of its neighbor nodes, and then weighted by smoothing factors.
      Thus, the amount of data exchange is cut-down to the halo grids.
   \item Only after a certain distance into the halo, the weighting factors
      starts to taper the remaining tail of the halo down to zero.  This
      leaves the immediate neighboring grids unchanged and preserves the
      derivatives on the core grids.  In the meantime the FFT on local
      data evades wraparound effect.
   \item The FFT and its inverse are performed on local data only, as shown in
      Fig.~\ref{fig:overlapping}.  The $N_I$ local grid data consists of the
      left halo $0,\ldots,i_0-1$, the right halo $i_1+1,\ldots,N_I-1$, and the
      internal core grids $i_0,\ldots,i_1$.  Only the core grids are updated at
      each time-marching.  Going to 3D, the $N_{\bar{x}}$, $N_{\bar{y}}$ and
      $N_{\bar{z}}$ in Eq.~(\ref{eq:PSTDupdating}) are then the number of
      grids in the local sub-domain.
\end{itemize}

In our implementation, Eq.~(\ref{eq:xi}) also serves as the smooth tapering
function.  Let $n_\text{t}$ be the number of grids for tapering.  We require the
halo width $n_\text{halo}$ be at least $n_\text{t}+4+1$.  This guarantees at
least 4 proximal grids with unchanged data and makes the derivatives more
accurate than the 8th-order central finite difference stencil.  $n_\text{t}$ is
a number to control the steepness of tapering.  The extra one is reserved for
holding 0 to overcome FFT wraparound.  Then the thickness for data exchange is
$n_\text{t}+4$.  Now the weighting factors are given as
\begin{equation}
   w[l]=\xi\left(\frac{l}{n_\text{t}+1}\right),\quad l=1,\ldots,n_\text{t}.
   \label{eq:weight}
\end{equation}
The index $i$ is counted longitudinally from distal to proximal, and only
$n_\text{t}$ layers of grids are weighted.

\section{Near-to-Distant-Field Transformation}
To both the Maxwell equations and the Schr\"{o}dinger equation, it is
challenging to obtain numerical solutions in infinite domain from finite domain
computations.  Scattering problems utilize the solutions at remote detection
sites, in either the Fresnel region or the far field.  It would be unwise to
create lattice to cover the entire space from the interaction region all the way
to the detection point.  The vast grid number makes this problem intractable,
costly and impractical.  FDTD\slash PSTD solvers of the Maxwell equations employ
the near-to-far-field transformation (NTFF), a direct outcome of the surface
equivalence theorem of electromagnetic fields, to obtain far-field
solution~\cite{book:Taflove2005}.  In
NTFF, an imaginary virtual box enclosing the entire interaction region is set.
By the surface equivalence theorem, the original interaction structure can be
replaced by the surface currents on the box \textit{with the inside structure
totally nulled out}.  The surface currents come from the direct FDTD/PSTD
computations in the internal model, and field values outside the box become
surface integrals of these currents with the Green's function.  Integrals on
merely the six surfaces of the box significantly relieve the computation burden.
Below, we derive the quantum version of the theorem, and identify the surface
terms required for the integral.  Because the Fresnel region is also concerned,
the word near-to-distant-field (NTDF) is more appropriate than NTFF.
Furthermore, in order to maintain numerical accuracy in the Fresnel region, a
semi-analytical approach is designed to conduct the surface integrals.

\subsection{Surface Equivalence Theorem}
Similar to the derivation of the surface equivalence theorem for electromagnetic
fields\cite{article:Lindell1996}, a enclosed surface $S$ is first defined to
completely surround the interaction area of the Schr\"{o}dinger potential.  This
implies the potential has finite range.  If the potential is unbounded in range,
a range cutoff is valid if the potential drops rapidly as $r$ increases.

Let the volume bounded by $S$ be $V$.  A step function is defined on $V$:
\begin{equation}
   P(\mathbf{r})=
   \begin{cases}
      0 & \text{if $\mathbf{r}\in V$,}\\
      1 & \text{if $\mathbf{r}\notin V$.}
   \end{cases}
   \label{eq:P}
\end{equation}
Note this definition is opposite to Ref.~\cite{article:Lindell1996} because our
goal is the solution outside $V$.  The derivative of a step function equals the
delta function, so the gradient of $P(\mathbf{r})$ would be nontrivial only at
the interface $S$, but $0$ elsewhere.  We can symbolically write
\begin{equation}
   \nabla P(\mathbf{r})=\mathbf{n}(\mathbf{r})\,\delta_S(\mathbf{r})
   \label{eq:gradP}
\end{equation}
where $\mathbf{n}$ is the outward surface normal of $V$.  Any volume integral
involving the surface delta function $\delta_S(\mathbf{r})$ would reduce to a
surface integral over $S$.

Consider the time-independent Schr\"{o}dinger equation and its Green's function, i.e.,
\begin{eqnarray}
	-\frac{\hbar^2}{2m}\nabla^2\Psi(\mathbf{r})+V(\mathbf{r})\Psi(\mathbf{r})=E\Psi(\mathbf{r}),\\
	\nabla^2G(\mathbf{r}|\mathbf{r_0})+k^2G(\mathbf{r}|\mathbf{r_0})=\delta(\mathbf{r}-\mathbf{r_0})
\end{eqnarray}
with $E$ the energy of the incident plane wave, $\mathbf{k}$ the wave vector,
and $k=\sqrt{2mE}/\hbar$.  Define a new wave function
\begin{equation}
   \Psi_\text{DF}(\mathbf{r})\equiv P(\mathbf{r})\Psi(\mathbf{r}).
   \label{eq:PsiDF}
\end{equation}
Here, $\Psi_\text{DF}$ is identical to $\Psi$ in the distant field, but vanishes
inside $V$.  In addition, the potential term $P(\mathbf{r})V(\mathbf{r})=0$.
This leads to the following,
\begin{equation}
   \nabla^2\Psi_\text{DF}(\mathbf{r})+k^2\Psi_\text{DF}(\mathbf{r})
   =\left(\nabla^2P(\mathbf{r})+2\nabla
   P(\mathbf{r})\cdot\nabla\right)\Psi(\mathbf{r}).
   \label{eq:PsiDFsurfpotential}
\end{equation}
As both $\nabla^2P$ and $\nabla P$ are non-zero only on $S$, the r.h.s of
Eq.~(\ref{eq:PsiDFsurfpotential}) serves as the surface source for
$\psi_\text{DF}$.  All the internal details of $V(\mathbf{r})$ do not explicitly
appear in Eq.~(\ref{eq:PsiDFsurfpotential}).  However, $V(\mathbf{r})$
implicitly determines the distant field wave $\Psi_\text{DF}$ through the
near-field waves on $S$, i.e. $\Psi$ and $\nabla\Psi$, which are solved in the
internal model.

A straightforward calculation presents
\begin{equation}
   \Psi_\text{DF}(\mathbf{r})=\oiint_S d\mathbf{s^\prime}\cdot\left[
      G(\mathbf{r^\prime}|\mathbf{r})\,\nabla^\prime\Psi(\mathbf{r^\prime})-
      \nabla^\prime G(\mathbf{r^\prime}|\mathbf{r})\,\Psi(\mathbf{r^\prime})\right],
   \label{eq:surfequivtheorem}
\end{equation}
where $d\mathbf{s^\prime}$ is the surface element of outward unit normal.
Using the Green's function expression
\begin{equation}
   G(\mathbf{r^\prime}|\mathbf{r})=-\frac{e^{ik|\mathbf{r^\prime}-\mathbf{r}|}}
   {|\mathbf{r^\prime}-\mathbf{r}|},
   \label{eq:G}
\end{equation}
Eq.~(\ref{eq:surfequivtheorem}) is further simplified to
\begin{equation}
   \Psi_\text{DF}(\mathbf{r})=-\frac{1}{4\pi}\oiint_S d\mathbf{s^\prime}\cdot
   \Biggl[\nabla^\prime\Psi(\mathbf{r'})
   +\left(-ik+\frac{1}
   {|\mathbf{r^\prime}-\mathbf{r}|}\right)\frac{\mathbf{r^\prime}-\mathbf{r}}
   {|\mathbf{r^\prime}-\mathbf{r}|}\Psi(\mathbf{r^\prime})\Biggr]
   \frac{e^{ik|\mathbf{r^\prime}-\mathbf{r}|)}}{|\mathbf{r^\prime}-\mathbf{r}|}.
   \label{eq:NTDF}
\end{equation}

So far in the derivation, the $\Psi$ in Eq.~(\ref{eq:surfequivtheorem}) are the
total wave function.  Recall that the virtual surfaces are located in free
space.  For scattering problems, the incident plane wave $\psi^\text{inc}$ is a
solution to the Schr\"{o}dinger equation in infinite free space.  By setting the potential
$V(\mathbf{r})=0$, as a special case of the above proof, $\psi^\text{inc}$
independently satisfies Eq.~(\ref{eq:surfequivtheorem}).  Since
$\Psi^\text{total}=\Psi^\text{scat}+\psi^\text{inc}$, the scattered wave function
$\Psi^\text{scat}$ also obeys Eq.~(\ref{eq:surfequivtheorem}).

\begin{figure}
   \centering
   \includegraphics[width=0.8\linewidth]{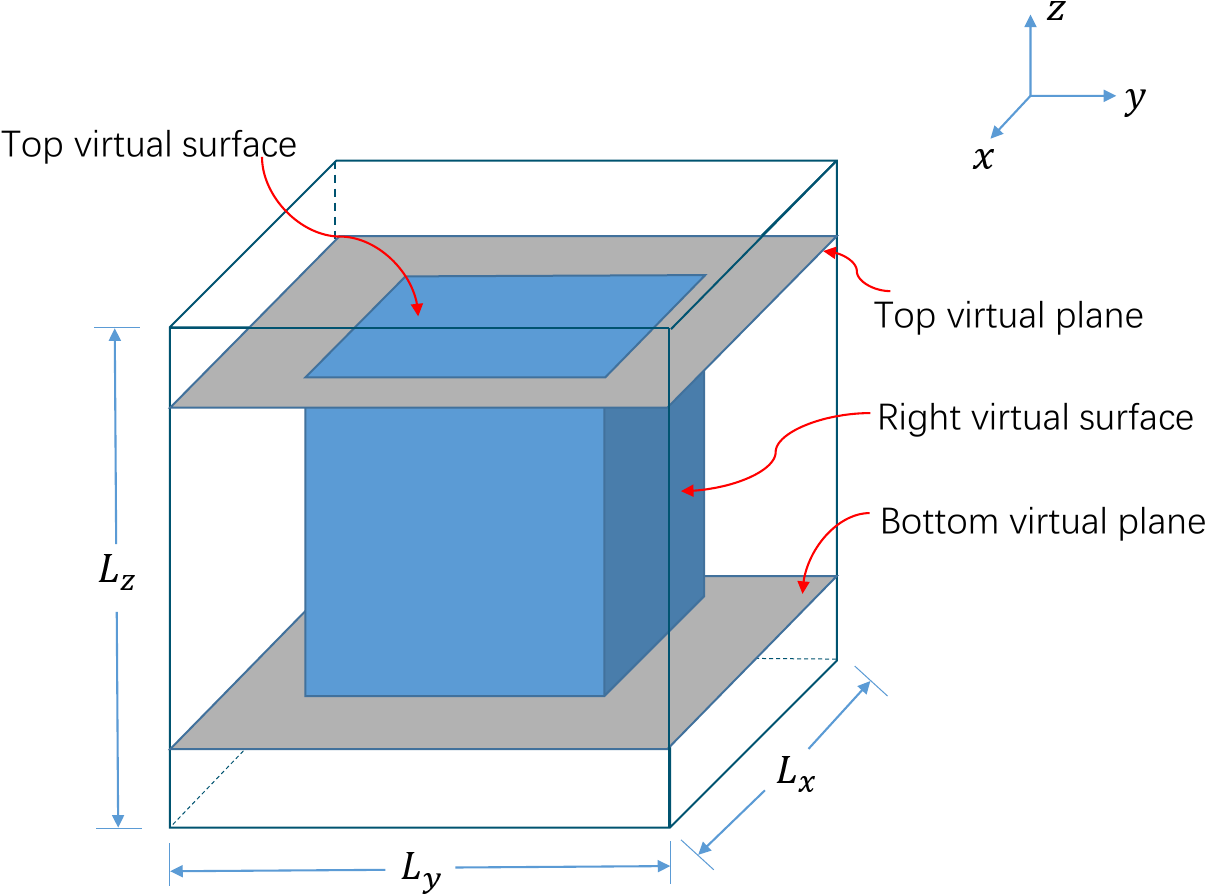}
   \caption{The virtual surface is part of the virtual plane which is an
	cross section of the internal model (Fig.~\ref{fig:model-setup}.)}
   \label{fig:virtual-surface}
\end{figure}

The surface equivalence theorem (Eq.~(\ref{eq:surfequivtheorem})) makes
processing quantum scattering of arbitrary potential possible, whenever the
potential goes to $0$ fast enough as the distance $r$ increases.  This only
requirement on potential over $r$ is the same as the conventional partial-wave
method on central potentials.  There is no other restrictions  on the detailed
form of $V(\mathbf{r})$ or on the specific shape of $S$.  Therefore, scattering by
a central or non-central potential, of scalar form or vector-dependent, or
spin-magnetic-field interactions can all be solved.  In practice, $S$ is often
arranged as a virtual rectangular box in the near field region of the scattered
field.  Thus, the integral of Eq.~(\ref{eq:NTDF}) simplifies to integration on
its six surfaces, and the gradient operation $\nabla^\prime$ on each surface
becomes one single derivate, for example, on $x$ for the front/back surfaces.
Finally, Eq.~(\ref{eq:NTDF}) identifies the surface terms required for the NTDF
transformation be the $\Psi$ and $\nabla\Psi$ on the six virtual surfaces.

However, in actual modeling, surface terms on slightly larger planes are stored,
in order to facilitate the 2D-FFT described in the next section.  In
Fig.~\ref{fig:virtual-surface}, the shaded box represents the virtual enclosure
of size $L^v_x\times L^v_y\times L^v_z$.  Its surfaces are referred to as the
vitual surfaces.  Each virtual surface sits on a virtual plane, which is a body
cross section of the internal model.  During the FDTD/PSTD time iterations, it
is the $\Psi$ and $\nabla\Psi$ on the six virtual planes that will be
accumulated and stored.  Due to the absorbing boundary
(Fig.~\ref{fig:model-setup}), $\Psi$ and $\nabla\Psi$ decay to $0$ at the edges.
Therefore, the 2D-FFT on the virtual planes (not the virtual surfaces) can
automatically avoid the wrap-around effect.

\subsection{Semi-Analytical Integration for NTDF}
FDTD electrodynamics applies a far-field approximation to simplify the surface
integral, i.e.,
\begin{equation}
   \frac{e^{ik|\mathbf{r}-\mathbf{r^\prime}|}}
   {|\mathbf{r}-\mathbf{r^\prime}|}
   \approx\frac{e^{ikr}}{r}
   e^{-ikr^\prime\cos\varPhi}
   \label{eq:FDTDfarfieldapprox}\\
\end{equation}
where $\mathbf{r}$ is the far field location, $\mathbf{r^\prime}$ the surface
grid, and $\varPhi$ the angle between $\mathbf{r}$ and $\mathbf{r^\prime}$.
Assume the virtural box size is $L$.  The far field condition $kL^2\ll r$ is
violated in the Fresnel region, rendering the phase expansion invalid.  In
addition, the denominator on the l.h.s of Eq.~(\ref{eq:FDTDfarfieldapprox})
$|\mathbf{r}-\mathbf{r^\prime}|$ also significantly deviates from $r$.
Fortunately, the FDTD\slash PSTD grids are typically of subwavelength size.  The
Fresnel field does satisfy the far-field condition of each surface cell.  The
denominator $|\mathbf{r}-\mathbf{r^\prime}|$ can be replaced with the distance
between the field coordinate and the cell center, and thus will be different for
different cells.  The overall surface integral (Eq.~(\ref{eq:NTDF})) can then be
disassembled into summations of the integrals on each individual cells.

\begin{figure}
   \centering
   \includegraphics[width=0.8\linewidth]{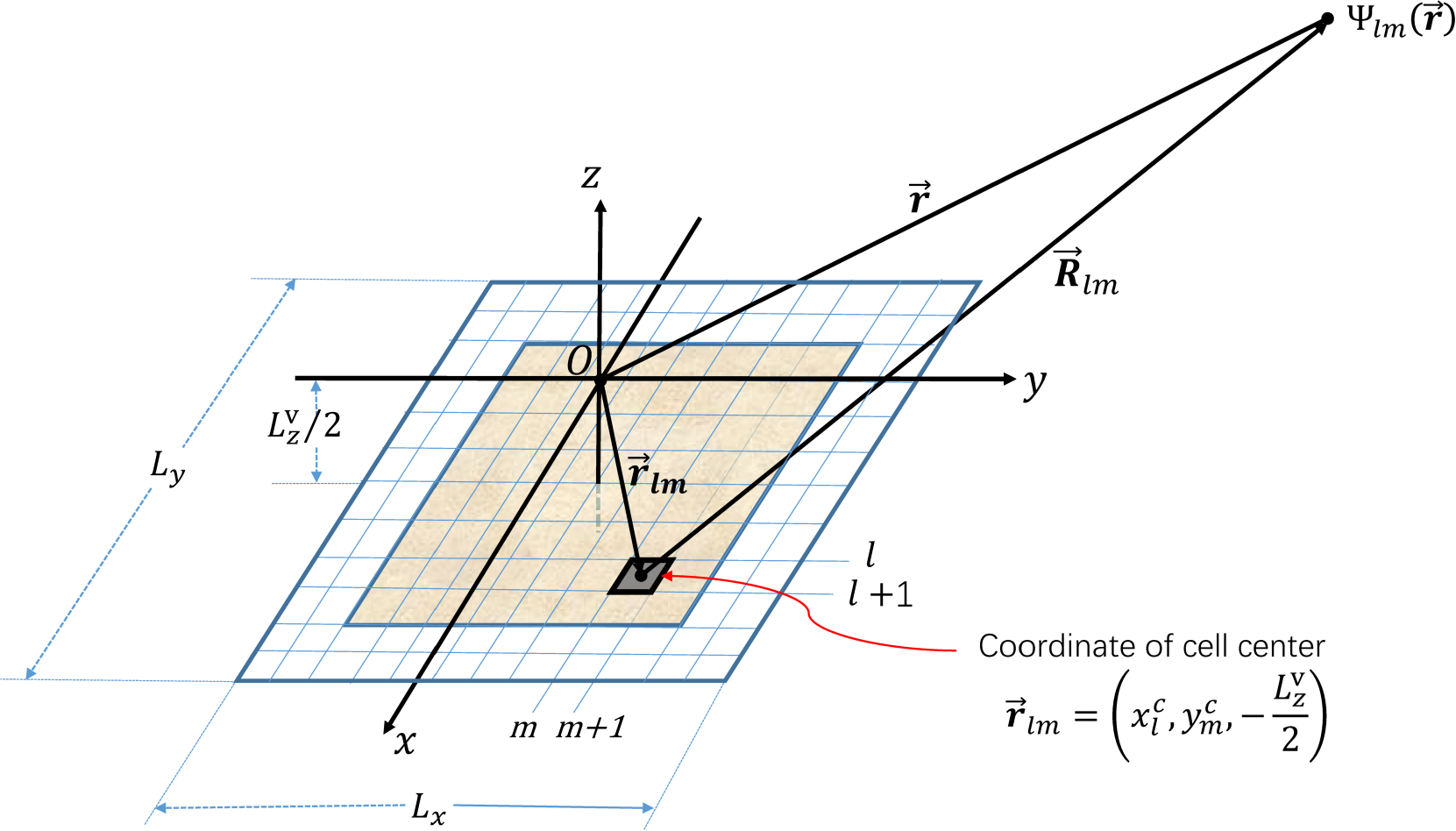}
   \caption{An illustration of the integration on a cell $(l,m)$ at the bottom virtual
   surface.}
   \label{fig:NTDF}
\end{figure}

A mathematical technique can further improve the numerical accuracy of the
integration, especially useful for PSTD, since PSTD typically sets coarser grids
than FDTD.  To reformat a discrete function into an analytical expression, first
we conduct FFT on the discrete data, then we use the results as the coeffients
for the inverse Fourier transform.  Without losing generality, we describe the
procedure on the bottom surface,
\begin{eqnarray}
   \psi^\text{b}_\text{an}(x,y)
   &=&\frac{1}{N_xN_y}\sum_{i^\prime=-N_x/2}^{N_x/2-1}
   \sum_{j^\prime=-N_y/2}^{N_y/2-1}
   \left.\mathcal{F}_{xy}\left[\psi^\text{b}\right]\right|
   _{i^\prime j^\prime}\nonumber\\
   &&\quad\exp\left[i\frac{2\pi i^\prime}{N_x\Delta x}
   \left(x+\frac{L_x}{2}\right)+i\frac{2\pi j^\prime}
   {N_y\Delta y}\left(y+\frac{L_y}{2}\right)\right],
   \label{eq:AnBottomPsi}
\end{eqnarray}
where $\mathcal{F}_{xy}$ is the 2D-FFT on the discrete $\psi^\text{b}$, $L_x$ and
$L_y$ are the size of the bottom virtual plane (Fig.~\ref{fig:NTDF}).
Expression for $\partial\psi^\text{b}/\partial z$ is similar.

Eq.~(\ref{eq:AnBottomPsi}) allows the integral on a surface cell to be
carried out semi-analytically.  Figure~\ref{fig:NTDF} illustrates cell
$(l,m)$ on the bottom virtual surface.  Let $x^c_l=(x_l+x_{l+1})/2$,
$y^c_m=(y_m+y_{m+1})/2$.  The coordinate of the cell center is
$\mathbf{r}_{l m}=\left(x^c_l,y^c_m,-L^v_z/2\right)$, $\mathbf{R}_{l
m}=\mathbf{r}-\mathbf{r}_{lm}$, and unit vector $\mathbf{\hat{R}}_{l
m}=\mathbf{R}_{lm}/R_{lm}$.  Within this cell, an off-cell-center location
$\mathbf{r^\prime}$ satisfies
$\mathbf{r}-\mathbf{r^\prime}=\mathbf{R}_{lm}-\left(\mathbf{r^\prime}
-\mathbf{r}_{lm}\right)$, where $\delta\mathbf{r^\prime}=\mathbf{r^\prime}
-\mathbf{r}_{lm}$ is the local offset from the center and coplanar with  the
cell surface.  Because $\Delta x$ and $\Delta y$ are subwavelength size, the
Fresnel region always satisfies $(\Delta x)^2\ll \lambda R$.  The
Green's function (Eq.~(\ref{eq:G})) can now be approximated as
$-\frac{e^{ikR_{lm}}}{R_{lm}}e^{-ik\mathbf{\hat{R}}_{lm}\cdot\mathbf{\delta
r^\prime}}$.
Consequently, the contribution from bottom surface cell $(l,m)$ to the
wave function at $\mathbf{r}$ is
\begin{eqnarray}
   \Psi^\text{b}_{lm}(\mathbf{r})
   &=&\frac{\Delta x\Delta y e^{ikR_{lm}}}{4\pi N_xN_yR_{lm}}
   \sum_{i^\prime=-N_x/2}^{N_x/2-1}\sum_{j^\prime=-N_y/2}^{N_y/2-1}
   (-1)^{i^\prime+j^\prime} e^{2\pi
   i\left(i^\prime\frac{x^c_l}{L_x}+j^\prime\frac{y^c_m}{L_y}\right)}.\nonumber\\
   &&\quad\left\{\left.\mathcal{F}_{xy}\left[\frac{\partial\psi}{\partial
   z}\right]\right|_{i^\prime j^\prime}
   +\left(ik-\frac{1}{R_{lm}}\right)\mathbf{\hat R}_{lm}\cdot\mathbf{\hat
   z}\left. \mathcal{F}_{xy}\left[\psi\right]\right|_{i^\prime j^\prime}\right\}\nonumber\\
   &&\quad\sinc\left(\pi\frac{i^\prime}{N_x}-\frac{k\Delta x}{2}
   \mathbf{\hat{R}}_{lm}\cdot\mathbf{\hat{x}}\right)
   \sinc\left(\pi\frac{j^\prime}{N_y}-\frac{k\Delta
   y}{2}\mathbf{\hat{R}}_{lm}\cdot\mathbf{\hat{y}}\right).
   \label{eq:bottomPsi_lm}
\end{eqnarray}
On the r.h.s of the above equation, we have suppressed the "bottom" superscript
to the $\psi$ for clarity.  Similarly, the contribution from the top surface
cell at $\mathbf{r}_{l m}=\left(x^c_l,y^c_m,L^v_z/2\right)$ is
\begin{eqnarray}
   \Psi^\text{t}_{lm}(\mathbf{r})
   &=&-\frac{\Delta x\Delta y e^{ikR_{lm}}}{4\pi N_xN_yR_{lm}}
   \sum_{i^\prime=-N_x/2}^{N_x/2-1}\sum_{j^\prime=-N_y/2}^{N_y/2-1}
   (-1)^{i^\prime+j^\prime} e^{2\pi
   i\left(i^\prime\frac{x^c_l}{L_x}+j^\prime\frac{y^c_m}{L_y}\right)}\nonumber\\
   &&\quad\left\{\left.\mathcal{F}_{xy}\left[\frac{\partial\psi}{\partial
   z}\right]\right|_{i^\prime j^\prime}+\left(ik-\frac{1}{R_{lm}}\right)\mathbf{\hat R}_{l m}\cdot\mathbf{\hat
   z}\left.\mathcal{F}_{xy}\left[\psi\right]\right|_{i^\prime j^\prime}\right\}\nonumber\\
   &&\quad\sinc\left(\pi\frac{i^\prime}{N_x}-\frac{k\Delta x}{2}\mathbf{\hat{R}}_{lm}\cdot\mathbf{\hat{x}}\right)
   \sinc\left(\pi\frac{j^\prime}{N_y}-\frac{k\Delta
   y}{2}\mathbf{\hat{R}}_{lm}\cdot\mathbf{\hat{y}}\right).
   \label{eq:topPsi_lm}
\end{eqnarray}

For a back\slash front surface cell,
\begin{eqnarray}
   \Psi^\text{k/f}_{mn}(\mathbf{r})
   &=&\pm\frac{\Delta y\Delta z e^{ikR_{mn}}}{4\pi N_yN_zR_{mn}}
   \sum_{j^\prime=-N_y/2}^{N_y/2-1}\sum_{k^\prime=-N_z/2}^{N_z/2-1}
   (-1)^{j^\prime+k^\prime} e^{2\pi
   i\left(j^\prime\frac{y^c_m}{L_y}+k^\prime\frac{z^c_n}{L_z}\right)}\nonumber\\
   &&\quad\left\{\left.\mathcal{F}_{yz}\left[\frac{\partial\psi}{\partial
   x}\right]\right|_{j^\prime k^\prime}+\left(ik-\frac{1}{R_{mn}}\right)
   \mathbf{\hat R}_{mn}\cdot\mathbf{\hat
   x}\left.\mathcal{F}_{yz}\left[\psi\right]\right|_{j^\prime k^\prime}\right\}\nonumber\\
   &&\quad\sinc\left(\pi\frac{j^\prime}{N_y}-\frac{k\Delta y}{2}\mathbf{\hat{R}}_{mn}
   \cdot\mathbf{\hat{y}}\right)\sinc\left(\pi\frac{k^\prime}{N_z}
   -\frac{k\Delta z}{2}\mathbf{\hat{R}}_{mn}\cdot\mathbf{\hat{z}}\right)
   \label{eq:backfrontPsi_mn}
\end{eqnarray}
with $+/-$ for the back\slash front surfaces, respectively.  The definitions of
$R_{mn}$, $y^c_{m}$ and $z^c_{n}$ are similar to the bottom surface.

For a left\slash right surface cell, we have
\begin{eqnarray}
   \Psi^\text{l/r}_{ln}(\mathbf{r})
   &=&\pm\frac{\Delta x\Delta z e^{ikR_{ln}}}{4\pi N_xN_zR_{ln}}
   \sum_{i^\prime=-N_x/2}^{N_x/2-1}\sum_{k^\prime=-N_z/2}^{N_z/2-1}
   (-1)^{i^\prime+k^\prime} e^{2\pi
   i\left(i^\prime\frac{x^c_l}{L_x}+k^\prime\frac{z^c_n}{L_z}\right)}\nonumber\\
   &&\quad\left\{\left.\mathcal{F}_{xz}\left[\frac{\partial\psi}{\partial
   y}\right]\right|_{i^\prime k^\prime}+\left(ik-\frac{1}{R_{ln}}\right)\mathbf{\hat R}_{ln}\cdot\mathbf{\hat
   y}\left.\mathcal{F}_{xz}\left[\psi\right]\right|_{i^\prime k^\prime}\right\}\nonumber\\
   &&\quad\sinc\left(\pi\frac{i^\prime}{N_x}-\frac{k\Delta x}{2}\mathbf{\hat{R}}_{ln}\cdot\mathbf{\hat{x}}\right)
   \sinc\left(\pi\frac{k^\prime}{N_z}-\frac{k\Delta z}{2}\mathbf{\hat{R}}_{ln}\cdot\mathbf{\hat{z}}\right)
   \label{eq:leftrightPsi_ln}
\end{eqnarray}
with $+/-$ for the left\slash right surfaces, respectively.

Finally, the scattered wave function at the distant location $\mathbf{r}$ is
\begin{equation}
   \Psi(\mathbf{r})=\sum_{l,m}\left[\Psi^\text{b}_{lm}(\mathbf{r})+\Psi^\text{t}_{lm}(\mathbf{r})\right]
   +\sum_{m,n}\left[\Psi^\text{k}_{mn}(\mathbf{r})+\Psi^\text{f}_{mn}(\mathbf{r})\right]
   +\sum_{l,n}\left[\Psi^\text{l}_{ln}(\mathbf{r})+\Psi^\text{r}_{ln}(\mathbf{r})\right].
   \label{eq:Psi}
\end{equation}
The summation indices $l$, $m$ and $n$ are within the virtual surfaces.

\subsection{Extracting Virtual Surface Data from Time-Domain Computations}
The inputs to the r.h.s of
Eqs.~(\ref{eq:bottomPsi_lm})-(\ref{eq:leftrightPsi_ln}) are all
time-independent.  These are actually the phasors of the time-dependent wave
function and its derivatives.  On the other hand, the outputs of
FDTD\slash PSTD internal model calculations are time-marching values.  Section~8.3 of
Ref.~\cite{book:Taflove2005} prescribes a procedure to extract phasor quantities
from time-varying fields.  During an FDTD\slash PSTD run, a recursive discrete
temporal Fourier transform is applied to the virtual surface fields "on the fly"
for each frequency of interest.  Such algorithm also works unchanged in the
Schr\"{o}dinger situation, i.e.
\begin{eqnarray}
   \left.\breve{\psi}\right|_{lm}&=&\sum_n\left.\psi\right|^n_{lm}e^{i\omega n\Delta t},
   \label{eq:psiphasor}\\
   \left.\left(\nabla\breve{\psi}\right)\right|_{lm}&=&\sum_n
   \left.\left(\nabla\psi\right)\right|^\text{n}_{lm}e^{i\omega
   n\Delta t}\label{eq:dpsiphasor},\\
   \left.\breve{\psi}^\text{inc}\right|_{O_xO_yO_z}&=&\sum_n\left.
   \psi^\text{inc}\right|^n_{O_xO_yO_z}e^{i\omega n\Delta t}.\label{eq:incpsiphasor}
\end{eqnarray}
Note the angular frequency $\omega$ actually means $\omega_1,\omega_2,\ldots$
for broadband incidence.  Eqs.~(\ref{eq:psiphasor})-(\ref{eq:incpsiphasor}) are
repeated for each frequency independently. 
Here, $(O_x,O_y,O_z)$ is the 1D origin where the incident wave makes
initial contact with the transition layer (Eq.~(\ref{eq:Oinc})).  The
stored phasor values which will be directly used in
Eqs.~(\ref{eq:bottomPsi_lm})-(\ref{eq:leftrightPsi_ln}) are the scaled ones,
\begin{eqnarray}
   \psi|_{lm}&=&\frac{\left.\breve{\psi}\right|_{lm}}
   {\left.\breve{\psi}^\text{inc}\right|_{O_xO_yO_z}},\label{eq:scaledphsiphasor}\\
   \left.\left(\nabla\psi\right)\right|_{lm}&=&
   \frac{\left.\left(\nabla\breve{\psi}\right)\right|_{lm}}
   {\left.\breve{\psi}^\text{inc}\right|_{O_xO_yO_z}}.\label{eq:scaleddpsiphasor}
\end{eqnarray}

This algorithm allows an impulsive wideband incident wave source condition.  The
response of the scattering system at multi-wavelengths can be obtained in one
run.  In comparison, a monochromatic sinusoidal wave source only provides the
response at one wavelength.  Multi-wavelength response will require multiple
runs, once for each wavelength.

For the case of sinusoidal wave incidence, we need to wait enough cycles for the
numerical results becoming steady.  Afterwards we start the temporal Fourier transform
(Eqs.~(\ref{eq:psiphasor})-(\ref{eq:incpsiphasor})) and continue the simulation
for one more cycle to accumulate the virtual surface phasor data.

For the case of pulsed incidence, the phasor accumulation starts at the first
time iteration and continues to the end.

\subsection{Validation of NTDF}
In order to test the accuracy of the NTDF transformation, we consider a
spherical wave originated from the center of a virtual box, generate the
required virtual surface data, input the data into 
Eqs.~(\ref{eq:AnBottomPsi})-(\ref{eq:Psi}), and check whether we can recover
the spherical wave function in the near field, the mid field (i.e., the
Fresnel region), and the far field.

In a parallel computing environment of two nodes, the internal model space is decomposed
into two subdomains, forming a $1\times 2\times 1$ topology.  A lattice of 
$384\times 192\times 384$ grids is built on each subdomain.  The overlapping halo is $15$
grids thick.  The spherical wave function $\exp(i\bar{r})/\bar{r}$ with $\bar{r}=kr$
is used as the golden standard.  The grid size is
$\Delta\bar{x}=\Delta\bar{y}=\Delta\bar{z}=\pi/10$, corresponding to 20 grids
per wavelength along the three axis directions.  The virtual surfaces are the
six planes $\bar{x}_\pm=\pm 41.3119$, $\bar{y}_\pm=\pm 36.5996$, and
$\bar{z}_\pm=\pm 41.3119$.  The distance from the origin to the diagonal corners is
68.9411.  This labels the near end of the external field.  The far field
condition, $\bar{r}_\text{threshold}\sim(\bar{x}_+-\bar{x}_-)^2/\lambdabar\approx 6800$,
characterizes the boundary between the Fresnel region and the far field.

\begin{figure}
   \centering
   \includegraphics[width=0.5\linewidth]{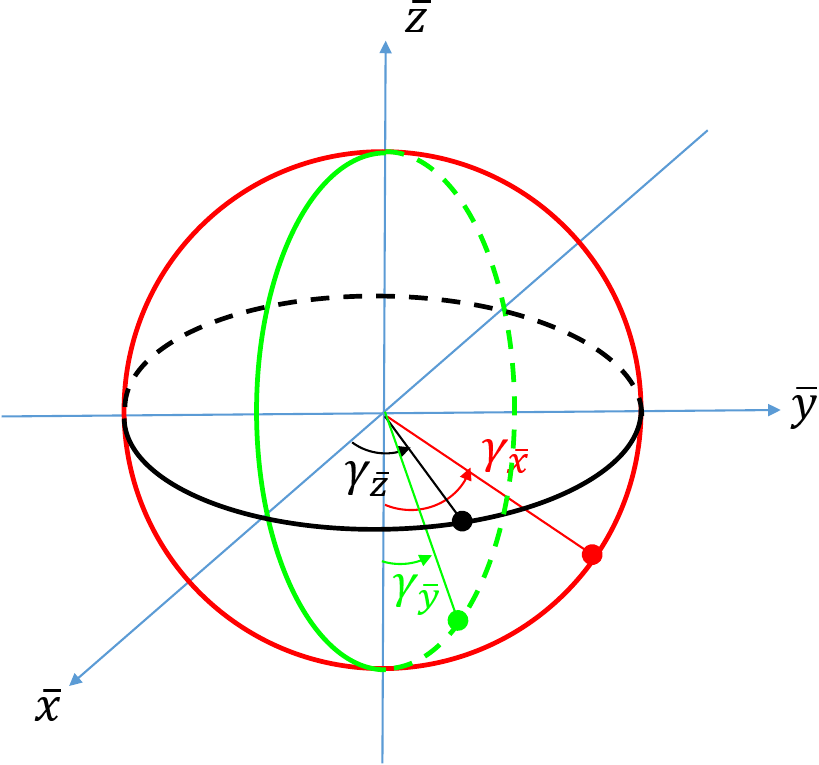}
   \caption{Definition of the Euler angle $\gamma$}
   \label{fig:Euler}
\end{figure}

\begin{figure}
   \centering
   \includegraphics[width=\linewidth]{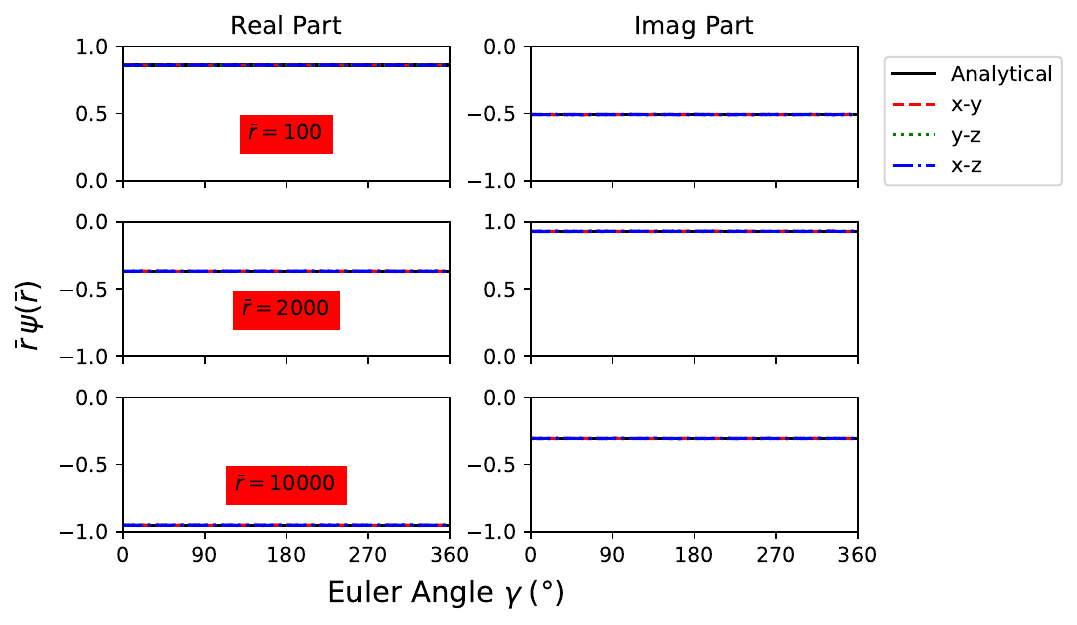}
   \caption{The reconstructed vs. the analytical spherical wave functions along
   the circles on the $x-y$, $y-z$, $x-z$ planes (Fig.~\ref{fig:Euler} at
   three different radii).}
   \label{fig:spherical}
\end{figure}

The surface terms of the spherical wave on the six virtual
surfaces are analytically generated.  The wave functions on three characteristic
planes (Fig.~\ref{fig:Euler}), the $x-y$, $y-z$, and $x-z$ planes, are
accordingly reconstructed.  The coordinate of
the wave function is best described by the Euler angles $(\alpha,\beta,\gamma)$
of the scattering plane (y-convention)~\cite{book:Sakurai2011}.
The first two Euler angle values $(\alpha,\beta)$ are: $x-y$ plane,
$(0^\circ,0^\circ)$; $y-z$ plane, $(0^\circ,90^\circ)$; $x-z$ plane,
$(90^\circ,90^\circ)$.  The definition of the Euler angle $\gamma$ is given in
Fig.~\ref{fig:Euler}.  Three radii, $\bar{r}=100,2000,10000$, are selected,
representing the near, mid, and far fields, respectively.
A spherical wave should be only a function of $\bar{r}$ and exhibit no
dependence on the three Euler angles.  In deed, the NTDF transformed wave
functions in Fig.~\ref{fig:spherical} show excellent agreement with 
the original analytical spherical wave at all external fields.
Furthermore, because the fields within the virtual surfaces have been
obtained by the direct FDTD\slash PSTD calculation, our technique can solve the quantum
potential scattering in the full 3D space.

\section{PSTD Validation Using the Scattering by Central Square Potentials}
A golden standard is required in order to validate our numerical algorithm.
Among existing scattering techniques, only the partial wave method on
spherically symmetric potential can provide relative rigorous numerical
solutions \textit{in the far field} after careful cutoffs of $l$ and the
potential range $r$~\cite{book:Thijssen2007}.  Other kind of methods or
other forms of potentials would need to
resort to some kind of approximations.  The partial wave method, the
decomposition of wave function into summations of $(l,m)$ components, and the
expression of the scattering cross-section in terms of the phase shifts
$\delta_l$ are the standard context of quantum mechanics
textbooks~\cite{book:Sakurai2011,book:Cohen-Tannoudji2005}.  Details in
numerically seeking $\delta_l$s are also well
described~\cite{book:Thijssen2007}.

\begin{figure}
   \centering
   \begin{subfigure}{0.45\linewidth}
      \includegraphics[width=\linewidth]{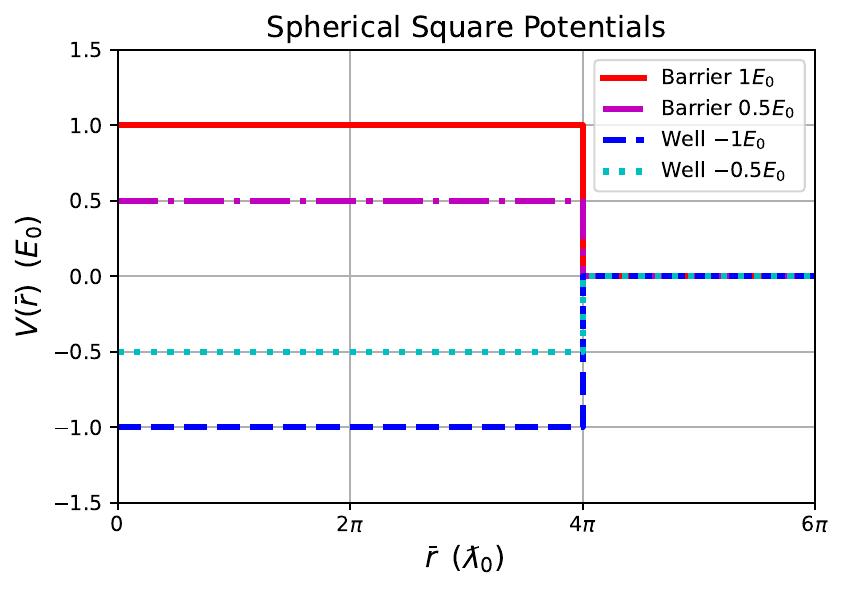}
      \caption{Central Square Potentials}
      \label{fig:subfig_potential}
   \end{subfigure}
   \hfill
   \begin{subfigure}{0.45\linewidth}
      \includegraphics[width=\linewidth]{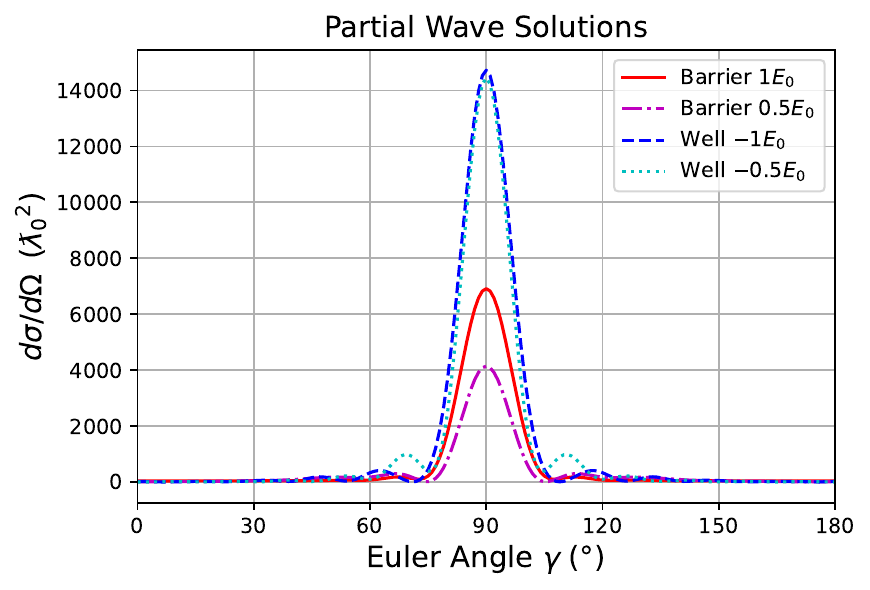}
      \caption{Partial Wave Cross-Sections}
      \label{fig:subfig_pwcrosssec}
   \end{subfigure}
   \caption{The partial wave solutions of four central square potentials.}
   \label{fig:subfigures}
\end{figure}

The central square potential for testing is given as
\begin{equation}
   V(\bar{r})=
   \begin{cases}
      s E_0 & \bar{r}\leqslant 4\pi\\
      0 & \bar{r}>4\pi
   \end{cases}
   \label{eq:sphericalV}
\end{equation}
with $E_0$ the energy of the incident sinusoidal plane wave,
$\bar{r}$($=r/\lambdabar_0)$ the reduced radius, and $s$ a scaling factor.  If
$s<0$, it is a potential well, and vice visa.  The potential's amplitude and
width are both characterized using the parameters of the incident wave.  In
Fig.~\ref{fig:subfig_potential} four central square potentials are set up, with
$s=1, 0.5, -0.5, -1$, respectively.  The radius is $4\pi$, corresponding to two
incident wavelengths.  The partial wave solutions of the far-field differential
cross-sections are given in Fig.~\ref{fig:subfig_pwcrosssec}.  Because the
incident direction is along the $y$-axis, in the $x-y$ and $y-z$ scattering
planes, the Euler angle $\gamma=90^\circ$ corresponds to the zero scattering
angle.

The four potentials in Fig.~\ref{fig:subfig_potential} are plugged into our
PSTD algorithm.  The size of the scatterer is $8\pi$,  so the threshold for the
far field would be $\bar{r}_\text{th}\approx 64\pi^2$.  The setup for the
modeling is as follows.  An overlapping domain decomposition of $1\times 2\times
1$ topology is configured on a two-node parallel platform.  A lattice
of $288\times 160\times 288$ grids is established on each subdomain, with
$\Delta\bar{x}=\Delta\bar{y}=\Delta\bar{z}=\pi/10$, corresponding to 20 grids
per wavelength.  The time increment is $\Delta\tau=\pi/1000$, per the
requirement of Eq.~(\ref{eq:dtau_stability}).  The widths of the ABC, the SF,
the transition layer, and the overlapping halo are 40, 41, 12, and 15 grids,
respectively.  The parameters for Eq.~(\ref{eq:Poshcl-Teller}) are $U_0=5.0$ and
$\alpha=0.1/\text{grid}$.  The virtual surfaces are set on the six middle planes
of the SF.  The sinusoidal incident wave is along the $y$-axis.

\begin{figure}
   \centering
   \includegraphics[width=\linewidth]{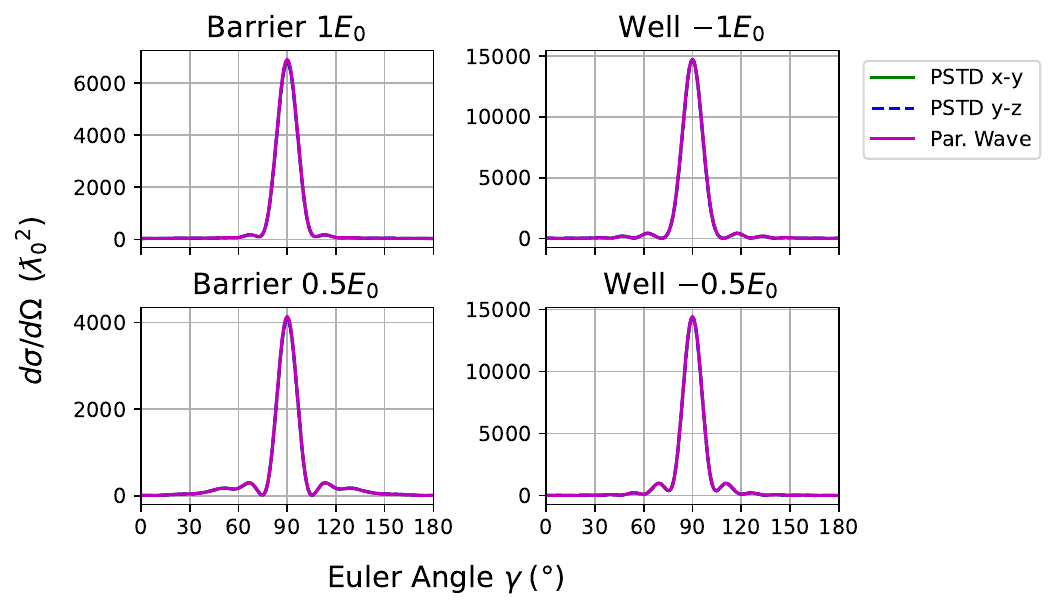}
	\caption{The differential cross-sections calculated using our algorithm
	at $\bar{r}=20000$ vs. the predictions by the partial wave method.}
   \label{fig:comparison}
\end{figure}

Figure~\ref{fig:comparison} presents our far field solutions to the scattering
cross-sections at $\bar{r}=2\times 10^4$, together with the partial wave
solutions in Fig.~\ref{fig:comparison}.  Clearly, the PSTD curves and the
partial wave predictions coincide in all cases.

\begin{figure}
   \centering
   \includegraphics[width=\linewidth]{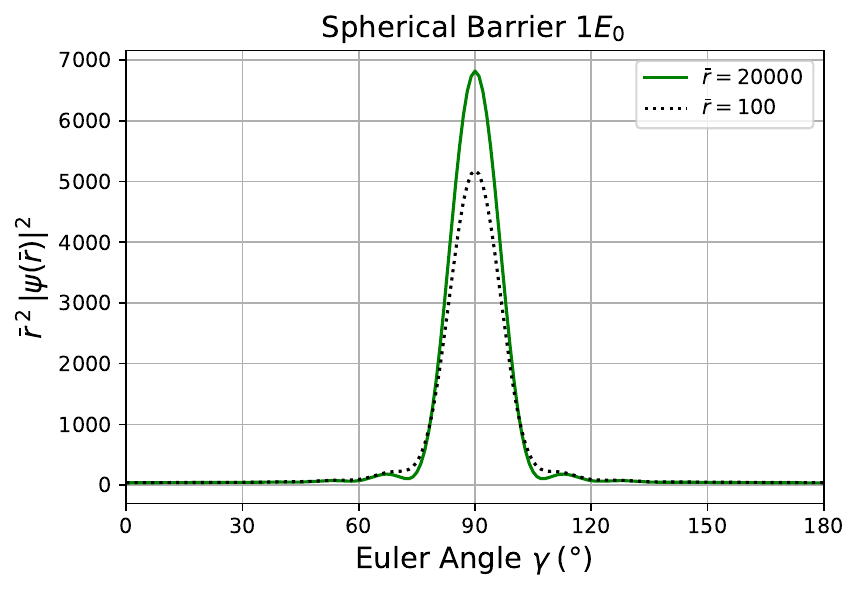}
   \caption{The Fresnel region vs. the far field.}
   \label{fig:Fresnel-Far}
\end{figure}

In the far field, the scatterer's size is viewed like a point, and thus the outgoing
scattering wave is reduced to a spherical wave.  However, in the near field and
the Fresnel region, the scatterer's size is nontrivial.  Set $\bar{r}=100$, a
distance in the Fresnel region, and acquire the wave function.
Figure~\ref{fig:Fresnel-Far} shows significant difference between the
Fresnel region and the far field.

\section{Conclusion}
In this work, we have developed a highly accurate numerical technique to solve the
problem of quantum potential scattering, by drawing analogy to the mature techniques of
computational electrodynamics.  The wave function in the interaction zone (i.e.
the internal field) and the close near field is directly solved by the
FDTD\slash PSTD
computations.  The surface terms on a virtual enclosing box are obtained.
The wave function in the free space outside the box is calculated through
the integration over these surfaces, derived from the surface equivalence
theorem.  The total-field\slash scattered-field scheme efficiently and accurately
introduces the incident wave into the model.  The TF enables the
processing of various potentials.  Except for the requirement of finite
force range, there is no other limit to the interaction forms, allowing
non-central forces, time-variant potentials, and nonlinear effects.

To validate the surface equivalence theorem, a spherical wave was precisely
reconstructed using NTDF, in regions from the near field, to the mid field
(i.e. Fresnel region), and to the far field.  The entire algorithm was tested
on four central square potentials.  The numerical results shew perfect agreement
with the partial wave predictions.  Further, a significant difference was
demonstrated between the Fresnel region and the
far field, indicating future simulation using Fresnel-region detection cannot
employ far-field results.  Immediately, the capability of processing arbitrary
potential scattering in the Fresnel region makes simulations on magnetic neutron
ghost imaging possible.  This work may see other potential applications in
studies of atomic and nuclear scatterings.

\section*{Acknowledgements}
This work was supported by the National Natural Science Foundation of
China under Grant Project No. 12075305.

\bibliographystyle{elsarticle-num}
\providecommand{\noopsort}[1]{}\providecommand{\singleletter}[1]{#1}%

\end{document}